\documentclass[journal]{IEEEtran}

\usepackage[OT1]{fontenc} 
\usepackage[numbers,sort&compress]{natbib}
\usepackage[cmex10]{amsmath}
\usepackage{amssymb}
\usepackage{bm}
\usepackage{braket}
\usepackage{graphicx}
\usepackage{color}
\usepackage{tabularx}
\usepackage{arydshln}
\usepackage{mathtools}
\usepackage{multirow}
\usepackage{algorithm,algorithmic}
\usepackage{url}
\usepackage{listings}
\usepackage{comment}
\usepackage{CJKutf8}
\usepackage[caption=false,font=footnotesize]{subfig} 
\usepackage{array}
\usepackage{booktabs}
\usepackage{tikz}
\usetikzlibrary{arrows.meta, positioning}

\def\CN{\mathcal{CN}}
\def\I{\mathbf{I}}

\def\H{\mathbf{H}}

\def\s{\mathbf{s}}

\def\y{\mathbf{y}}
\def\U{\mathbf{U}}
\def\C{\mathbb{C}}

\def\E{\mathbb{E}}
\def\S{\mathbf{S}}
\def\W{\mathbf{W}}
\def\F{\mathrm{F}}
\def\e{\mathrm{H}}

\def\G{\mathcal{G}}

\def\a{\mathbf{e}}
\def\w{\mathbf{w}}
\def\j{\mathrm{j}}

\begin{document}
\title{Low-Complexity and Power-Efficient Precoding Codebook Design on Sparse Grassmannian}

\author{Joe~Asano,~\IEEEmembership{Student Member,~IEEE}, Yuto Hama,~\IEEEmembership{Member, IEEE}, Hiroki Iimori,~\IEEEmembership{Member, IEEE},\\ Chandan Pradhan,~\IEEEmembership{Member, IEEE}, Szabolcs Malomsoky, and Naoki~Ishikawa,~\IEEEmembership{Senior~Member,~IEEE}.\thanks{J.~Asano and N.~Ishikawa are with the Faculty of Engineering, Yokohama National University, 240-8501 Kanagawa, Japan (e-mail: iskw@ieee.org). Y.~Hama, H.~Iimori, C.~Pradhan, and S.~Malomsoky are with Ericsson Research, Ericsson Japan K. K., Yokohama, 220-0012 Kanagawa, Japan (e-mail: [yuto.hama, hiroki.iimori, chandan.pradhan, szabolcs.malomsoky]@ericsson.com).
}}

\markboth{}
{Shell \MakeLowercase{\textit{et al.}}: Bare Demo of IEEEtran.cls for Journals}
\maketitle

\begin{abstract}
We propose a sparse Grassmannian design for precoding codebooks. Due to their sparse structure, our proposed codebooks achieve low peak-to-average power ratio~(PAPR), low complexity of precoder multiplication, and low storage cost, while demonstrating performance comparable to the optimal codebook.
Specifically, we introduce a method for constructing codebooks based on Schubert cell decomposition on the Grassmann manifold. Designing an optimal Grassmannian precoding codebook generally requires high computational complexity. In the proposed approach, by exploiting its sparsity, the objective function can be simplified, and the search space can also be significantly reduced compared to state-of-the-art codebooks. Numerical simulations in uplink systems demonstrate that the proposed sparse codebook asymptotically approaches the optimal codebook and outperforms the codebook currently adopted in 5G NR, in terms of achievable rate under uncorrelated Rayleigh fading channels, while maintaining substantially lower PAPR than conventional dense designs. These results confirm that the proposed sparse codebook can be a practical and power-efficient alternative to conventional codebooks for a wide range of uplink transmission scenarios.
\end{abstract}


\IEEEpeerreviewmaketitle

\section{Introduction}
\IEEEPARstart{T}{he} volume of wireless communication traffic continues to increase steadily, driven by the rapid expansion of real-time cloud services and ultra-high definition video applications~\cite{ericsson2025ericsson}. 
Although the current 5G standard~\cite{3gpp_ts_38211_v1820} achieves sub-millisecond latency and gigabit-per-second peak data rates through massive multiple-input multiple-output (MIMO) technology, the use of millimeter-wave spectrum, and advanced beam management, future beyond-5G systems aim to push performance even further. These next-generation networks target significant enhancements in ultra-high reliability, ultra-low latency, energy efficiency, spectral efficiency, and overall throughput~\cite{zhang20196G}. At the same time, the use of Internet of Things (IoT) and device-to-device communications is expected to expand further, requiring even more efficient systems with low complexity and low power consumption.
To address these challenges, sparsity plays a crucial role in modern wireless communications~\cite{ishikawa201850}. Due to its mathematical properties, sparsity enables low-complexity algorithms such as compressed sensing and orthogonal matching pursuit~\cite{choi2017compressed}, which are utilized in channel estimation~\cite{lee2016channel} and beamforming design~\cite{ayach2014spatially}. Sparsity is also used in spatial modulation schemes such as index modulation~\cite{ishikawa2016subcarrierindex}, providing low-complexity but high-performance transmission techniques. The effectiveness of sparsity has been demonstrated in the design of IoT networks~\cite{qin2018sparse}, and its future applications are highly anticipated.

In modern MIMO systems, precoding is an essential technique to achieve efficient array gain, and it is widely employed in current wireless standards. 
In 5G systems, both codebook-based and non-codebook-based precoding are supported. In particular, codebook-based precoding is widely used in practical deployments, in which the precoder is selected from predefined codebooks specified in~\cite{3gpp_ts_38211_v1820}.
For limited feedback MIMO systems, it has been shown that the optimal precoding codebook should be designed on the Grassmann manifold~\cite{love2003grassmannian, love2005limited, love2008overview}.
More broadly, the Grassmann manifold has also been well established as an effective constellation design for noncoherent communications~\cite{hochwald2000unitary, ngo2025noncoherent}. Motivated by this fundamental property, a large body of literature has investigated constellation design on the Grassmann manifold for noncoherent communications~\cite{hochwald2000systematic, kammoun2003new, kammoun2007noncoherent, gohary2009noncoherent, ngo2020cubesplit, ngo2022joint, cuevas2023union, cuevas2024constellations}.
The problem of maximizing the achievable rate under uncorrelated Rayleigh fading channels can be formulated as a Grassmann subspace-packing problem~\cite{conway1996packing}, which aims to maximize the minimum chordal distance (MCD) between codewords. However, as the number of transmit antennas and data streams increases, the precoder dimensions grow proportionally. This expansion enlarges the search space and the number of optimization variables, which is inconsistent with the demand for low-complexity and scalable designs. Thus, for practical and future precoding codebook designs, it is crucial to establish design guidelines for low-complexity yet high-performance codebooks, even under large-scale system parameters. 

Toward 6G, further enhancement of uplink capacity is expected.
In uplink systems, the high peak-to-average power ratio (PAPR) remains a major challenge, particularly when power-amplifier (PA) efficiency is critical, such as for cell-edge users.
Consequently, numerous studies have analyzed the statistical characteristics of PAPR~\cite{ochiai2001distribution, ochiai2012instantaneous, ochiai2013statistical} and proposed effective reduction techniques~\cite{taojiang2008overview, hama2023timefrequency, yao2019semidefinite} for orthogonal frequency-division multiplexing (OFDM) and discrete Fourier transform spread OFDM (DFT-s-OFDM), which are employed in the current 5G uplink systems. 
However, in 5G, DFT-s-OFDM is limited to single data stream transmission. For 6G, extending DFT-s-OFDM to support multiple data stream transmission is being discussed so as to improve UL capacity and coverage.
In practice, however, the PAPR of DFT-s-OFDM significantly depends on the precoder. In fact, the conventional precoders in 5G increase the PAPR of DFT-s-OFDM since existing codebooks employ dense precoder structures to maximize the performance in terms of array gain and are designed to support only single-stream DFT-s-OFDM, which breaks the single-carrier property.
Therefore, for the stream expansion of DFT-s-OFDM toward 6G, the precoder for this system must be carefully designed from both the PAPR and array gain perspectives.

Against this background, we propose a low-complexity and power-efficient precoding codebook design by introducing {sparsity}.
We introduce a method for designing the codebook based on the theory of Schubert cells on the Grassmann manifold~\cite{konishi2022novel, hansen2007schubert}. Furthermore, we evaluate the performance of the proposed sparse codebooks by comparing them with conventional methods, and we numerically demonstrate their effectiveness. The major contributions of the proposed method are summarized as follows:
\begin{enumerate}
    \item \textbf{Near-optimal performance}:
    We propose a novel sparse Grassmannian design for precoding codebooks via Schubert cell decomposition. We demonstrate that the proposed codebooks reduce both time and space complexity owing to sparsity. Numerical simulations show that the proposed codebooks achieve performance asymptotically approaching that of the optimal codebook, despite the sparsity constraint.
    \item \textbf{Reduced PAPR}:
    We numerically analyze the effect of precoder sparsity on PAPR in both OFDM and DFT-s-OFDM. The results indicate that the proposed sparse codebook significantly reduces PAPR in DFT-s-OFDM compared with conventional codebooks.
    \item \textbf{Scalable design}:
    For the codebook optimization problem of maximizing the MCD, the proposed codebook design method enables significant simplification of the objective function and a substantial reduction of the search space in a specific case. Consequently, the construction of codebooks requires only low-complexity optimization, even when the number of transmit antennas, data streams, and codebook size become large.
    Even with the simplified optimization process, the proposed codebooks demonstrate comparable performance to the optimal codebooks while maintaining flexibility in parameter design.
\end{enumerate}

The contributions of this paper are fundamentally different from those of our preprint~\cite{asano2026sparse}. While both papers share a common framework based on the sparse Grassmannian design, they address fundamentally different system models. Specifically, the previous work focuses on noncoherent communications, whereas the present paper considers precoding codebook design for limited feedback systems. This difference in system models leads to substantial differences in problem formulations, analytical developments, and resulting insights. Moreover, the constellation proposed in~\cite{asano2026sparse} is designed to minimize pairwise error probability. Its performance as a precoding codebook, as considered in this paper, {cannot be guaranteed}.

The remainder of this paper is organized as follows. Section~\ref{sec:grassmann} reviews the fundamentals of the Grassmann manifold and conventional codebooks. Section~\ref{sec:sys} presents the system model, and Section~\ref{sec:prop} introduces the proposed sparse Grassmannian design. Section~\ref{sec:analy} analyzes the characteristics of the proposed codebook, and Section~\ref{sec:res} provides numerical results. Finally, conclusions are drawn in Section~\ref{sec:conc}.

\section{Grassmannian Codebooks}
\label{sec:grassmann}
In this section, we introduce the Grassmann manifold and its applications to wireless communications. We first present the mathematical definition of the complex Grassmann manifold and later discuss several representative codebook construction methods.

\subsection{Mathematical Definition~\cite{bendokat2024grassmann, edelman1999geometry}}
For a positive integer $M$, the unitary group of order $M$, denoted by $\mathcal{U}(M)$, is defined as the set of all $M\times M$ complex matrices $\U$, which is defined as
\begin{align}
    \mathcal{U}(M)=\left\{\U\in \C^{M\times M} ~ | ~\U^{\e}\U=\I_{{M}} \right\}.
    \label{eq:unitary}
\end{align}
For $T>M$, the complex Stiefel manifold $\mathcal{S}(T,M)$ consists of all $T\times M$ matrices $\S$ with orthonormal columns, which is defined as
\begin{align}
    \mathcal{S}(T,M)=\left\{\S \in \C^{T\times M} ~ | ~\S^{\e}\S=\I_{{M}} \right\}.
    \label{eq:stiefel}
\end{align}
Each matrix $\S\in\mathcal{S}(T,M)$ represents an $M$-dimensional subspace of $\C^{T}$ spanned by its columns. Letting $\s_m$ denote the $m$th column of $\S$, the associated subspace, i.e., the linear span of $\S$ can be written as
\begin{align}
    \operatorname{span}(\S)=\left\{h_1\s_1+\cdots+h_{{M}}\s_{{M}}~|~h_1,\cdots,h_{{M}}\in\C \right\}.
    \label{eq:span}
\end{align}
Two matrices $\S_1,\S_2\in\mathcal{S}(T,M)$ are considered equivalent if they span the same subspace, that is, if $\text{span}(\S_1)=\text{span}(\S_2)$. The corresponding equivalence class of $\S$ is denoted by $[\S]$, which is written as 
\begin{align}
    [\S]=\left\{\S_1\in\mathcal{S}(T,M)~\mid~\S_1\sim\S_2 \right\}.
\label{eq:equivalence}
\end{align}
Then, the Grassmann manifold $\G(T,M)$ is defined as the set of all such equivalence classes and can be expressed as the quotient space by the unitary group $\mathcal{U}(M)$ of the Stiefel manifold $\mathcal{S}(T,M)$
\begin{align}
    \mathcal{G}(T,M)=\mathcal{S}(T,M) ~ / ~ \mathcal{U}(M)=\left\{ [\S]~|~\S\in\mathcal{S}(T,M) \right\}.
\end{align}

A key property of the Grassmann manifold is its invariance under right multiplication by unitary matrices. Specifically, for any $\W\in\G(T,M)$ and any $\U\in\mathcal{U}(M)$, the relation
\begin{align}
    \text{span}(\W)=\text{span}(\W\U),
    \label{eq:inv}
\end{align}
holds. This unitary invariance implies that different matrix representations corresponding to the same subspace are equivalent. Such a property is widely exploited in wireless communications, for example in noncoherent detection and channel estimation~\cite{ngo2025noncoherent,endo2024boosting, kato2025maximizing}. In this paper, a finite set of points on $\G(T,M)$ is referred to as a Grassmannian codebook.

Let $\mathcal{W}=\{\W_1,\cdots,\W_{|\mathcal{W}|}\}$ denote a Grassmannian codebook of size $|\mathcal{W}|$, where each codeword $\W_i$ represents a point on $\G(T,M)$. In this codebook, the chordal distance between two points $\W_i$ and $\W_j$, denoted by $d_{\mathrm{c}}(\W_i,\W_j)$ is defined as
\begin{align}
\begin{split}
    d_{\mathrm{c}}(\W_i,\W_j)&=\frac{\|\W_i\W_i^\e-\W_j\W_j^\e\|_\F} {\sqrt{2}} \\
    &=\sqrt{M-\|\W_i^\e\W_j\|_\F^2} 
    \label{eq:dch}.
\end{split}
\end{align}

\subsection{Conventional Grassmannian Codebooks} 
\subsubsection{Manopt~\cite{endo2024boosting, gohary2009noncoherent}}
The codebook design method based on direct optimization over the Grassmann manifold is referred to as Manopt in this paper. The method that maximizes the MCD, which is defined as
\begin{align}
    d_{\mathrm{min}} = \underset{1\leq i<j\leq |\mathcal{W}|} {\operatorname{min}} d_\mathrm{c}(\W_i,\W_j),
\end{align}
through optimization using (\ref{eq:dch}) is referred to as MCD-Manopt. The MCD maximization problem can be formulated as
\begin{align}
    \underset{\mathcal{W}}{\operatorname{maximize}}& \underset{1\leq i<j\leq |\mathcal{W}|} {\operatorname{min}} d_{\mathrm{c}}(\W_i,\W_j)
    \label{eq:obj_MCD}\\
    \text{s.t.}&~~\W_i\in\mathcal{G}(T,M),~\forall ~i=\left\{1,\cdots,|\mathcal{W}| \right\}, \nonumber
\end{align}
where the objective function in (\ref{eq:obj_MCD}) can be approximated by a smooth surrogate objective as
\begin{align}
    \underset{\mathcal{W}}{\operatorname{minimize}}~~&\log \underset{1\leq i<j\leq |\mathcal{W}|}{\sum}\exp\left(-\frac{\|\W_i\W_i^\e-\W_j\W_j^\e\|_\F}{\epsilon} \right)
    \label{eq:obj_MCD_2}\\
    \text{s.t.}~~&\W_i\in\mathcal{G}(T,M),~\forall ~i=\left\{1,\cdots,|\mathcal{W}| \right\}, \nonumber
\end{align}
with $\epsilon$ a smoothing constant. Based on the results from numerical optimization, MCD-Manopt can be regarded as the theoretically optimal codebook under uncorrelated Rayleigh fading channels~\cite{love2003grassmannian, love2005limited, love2008overview}.

\subsubsection{Exp-Map~\cite{kammoun2003new, kammoun2007noncoherent}}
Exponential map (Exp-Map) is one of the most structured design methods for the Grassmann manifold.
Using the matrix exponential, we can represent any point on the Grassmann manifold as
\begin{align}
    \W=\left[ \exp
    \begin{pmatrix}
        \mathbf{0} & \mathbf{\Theta} \\
        -\mathbf{\Theta}^\e & \mathbf{0} 
    \end{pmatrix}
     \right] \I_{T,M},
     \label{eq:exp}
\end{align}
where $\I_{T,M}=[\I_M ~\mathbf{0}_{M\times(T-M)}]^\top$ and $\mathbf{\Theta}\in\C^{M\times(T-M)}$ is an arbitrary matrix, such as space-time block codes composed of QAM symbols. {Exp-Map is a representative non-optimized codebook construction method that supports multi-stream scenarios, i.e., $M\geq2$. Although this method was originally proposed for noncoherent code design, it can also be used as a precoding codebook because its design guideline maximizes the MCD, and we adopt it as a benchmark in this paper.}

\subsubsection{5G NR Codebook~\cite{3gpp_ts_38211_v1820}}
The 5G NR codebook has been widely adopted in {current} 5G systems.
{Although this is not clearly specified in the standard,} 
each precoder corresponds to a point on the Grassmann manifold, and the corresponding codebooks are provided in tabular form according to the number of transmit antennas and data streams. 
While the specific {design} criteria are not described in the standard, it is assumed that the codebooks are designed based on DFT with antenna polarization.
{It is employed} as one of the benchmark schemes in our simulations, together with MCD-Manopt.
The specific NR codebooks used in our simulations are provided in {the} Appendix.

\section{System Model}
\label{sec:sys}
In this section, we describe the limited feedback MIMO precoding system~\cite{love2003grassmannian, love2005limited, love2008overview}, to which both conventional and proposed codebooks are applied. 

We consider a limited feedback MIMO system with $T$ transmit antennas and $N$ receive antennas. Let $M\leq \min(T,N)$ be the number of data streams. For a given channel matrix $\H\in\C^{N\times T}$, the perfect channel state information (CSI) is assumed to be available at the receiver in this system.
Let $\s \in \C^{M \times 1}$ denote the symbol vector, satisfying the average power constraint $\E[\s\s^\e]=\I_{M}$. The frequency-domain received signal $\y_k\in\C^{N\times 1}$ on subcarrier $k$ is given by
\begin{align}
    \y_k=\H_k\W_k\s_k+\mathbf{v}_k,
    \label{eq:sys_pre}
\end{align}
where $\W_k \in \mathbb{C}^{T \times M}$ is a precoding matrix and $\mathbf{v}_k \in \mathbb{C}^{N\times1}$ is the additive white Gaussian noise vector, with each entry independent and identically distributed as $\CN(0, \sigma_v^2)$.

In this system, we consider a frequency-flat fading channel and wideband precoding.
Both the channel and the precoder are assumed to remain constant across all subcarriers within a band. That is, $\H_k=\H$ and $\W_k=\W$ for all subcarriers $k$ in a band.
Under this assumption, we consider typical channel models such as Rayleigh fading and Rician fading~\cite{tse2005fundamentals}.
Each entry of the uncorrelated Rayleigh fading channel matrix $\H$ is independently and identically distributed as a complex Gaussian variable with zero mean and variance of one, i.e., $\CN(0,1)$. 
The Rician channel model is defined as a superposition of a line-of-sight (LoS) component~\cite{liu2015space, ishikawa2017generalized} and a non-line-of-sight (NLoS) component with Rician factor $K\ge0$, denoting the ratio of the LoS power to the NLoS power. 

The precoding matrix is selected from a finite codebook $\mathcal{W} = \left\{\W_1,\cdots,\W_{|\mathcal{W}|} \right\}$, which is shared in advance between the transmitter and receiver. Each matrix $\W_i$ satisfies the orthonormal constraint $\W_i^\e \W_i = \I_{M}$, and can thus be regarded as a point on the complex Grassmann manifold $\mathcal{G}(T,M)$. 
{At} that point, the constraint $M\le \min(T,N)$ must be satisfied because $M$ represents the number of data streams. Furthermore, this constraint automatically satisfies the conditions for the Grassmann manifold described in Section~\ref{sec:grassmann}-A.
Given perfect CSI, the receiver selects $\W_i\in\mathcal{W}$ that maximizes the achievable rate, which is denoted in~\cite{love2005limited} as
\begin{align}
    R(\W_i)=\log_2\det\left(\I_{M}+\frac{\rho}{M}\W_i^\e\H^\e\H\W_i \right),
    \label{eq:capacity_i}
\end{align}
where $\rho$ denotes the signal-to-noise ratio (SNR).
In this system, the average SNR at the receiver is defined as $\rho = 1/\sigma_v^2$.
The receiver then feeds back the index $\hat{i}$ corresponding to the optimal precoder matrix as
\begin{align}
   \hat{i}=\underset{1\leq i\leq|\mathcal{W}|}{\operatorname{argmax}}~R (\W_i), 
   \label{eq:selection}
\end{align}
in $\log_2|\mathcal{W}|$ bits. For limited feedback systems under uncorrelated Rayleigh fading channels, the optimal codebook design corresponds to maximizing the MCD of the codebook defined in (\ref{eq:obj_MCD}). Thus, the optimal precoding codebook is MCD-Manopt in terms of the achievable rate. 

The system model described above can be readily extended to OFDM and DFT-s-OFDM schemes employed in current 5G uplink transmission. In this extended system, {we apply optimal-index feedback via (\ref{eq:selection}) and precoding to each subcarrier symbol of the DFT-precoded sequence}. We then perform the oversampling procedure described in~\cite{berardinelli2008OFDMA, berardinelli2009feasibility} for accurate PAPR evaluation. 
In addition, we evaluate the PAPR of the time-domain signal at each transmit antenna. Let $x_t[n]$ denote the discrete-time baseband signal transmitted from the $t$-th antenna, then the PAPR for the $t$-th antenna is defined as 
\begin{align}
    \text{PAPR}(t)=\frac{\max_n|x_t[n]|^2}{\E\left[|x_t[n]|^2\right]},
    \label{eq:PAPR}
\end{align}
which represents the ratio of the maximum instantaneous power to the average transmit power.  
 
\section{Proposed Design Method}
\begin{figure*}[tb]
    \centering
    \includegraphics[width=\linewidth]{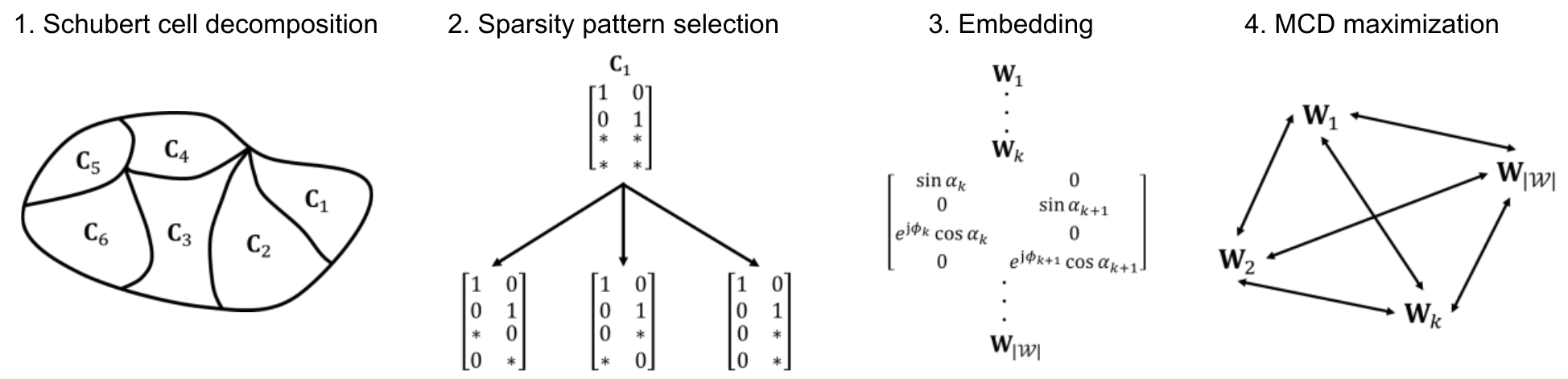}
    \caption{Overview of the proposed codebook design method that consists of four steps.}
    \label{fig:overview}
\end{figure*}

\label{sec:prop}
This section presents the design method of the proposed sparse Grassmannian codebook. The detailed steps of the design procedure are described for the general case, {followed by a specific example}.
\subsection{General Design Method}
The overall procedure can be summarized as shown in Fig.~\ref{fig:overview}.
The proposed method is applicable to any $T$ and $M$ that satisfy $T > M > 1$.
In contrast to conventional dense codebooks, this procedure systematically generates sparse structures, reducing the complexity associated with precoder multiplication, index selection, and storage cost.

\subsubsection{Schubert Cell Decomposition~\cite{konishi2022novel,hansen2007schubert}}
The complex Grassmann manifold $\mathcal{G}(T,M)$ can be decomposed into a disjoint union of Schubert cells
\begin{align}
    \mathcal{G}(T,M)=\bigsqcup_{\mathbf{k}\in\binom{[T]}{M}}\mathbf{C}_{\mathbf{k}},
    \label{eq:Schubert_decompose}
\end{align}
where each index $\mathbf{k} = \{k_1,\cdots,k_M\}$ defines a Schubert cell $\mathbf{C}_{\mathbf{k}}$, and the symbol $\bigsqcup$ denotes a disjoint union.
Each cell $\mathbf{C}_{\mathbf{k}}$ can be uniquely represented by a $T\times M$ matrix in column echelon form, such that the $(k_l,l)$-th entry is $1$ for $l=1,\cdots,M$, with all entries above that position in the same column and all entries to the left in the same row equal to zero. The other entries, i.e., to the right in the same row or below in the same column, may take arbitrary complex values. In the subsequent discussion in this paper, we denote this arbitrary complex number as $\ast$. 
This decomposition provides a structured parameterization of $\mathcal{G}(T,M)$ that naturally partitions the manifold into simpler subspaces. It enables us to construct sparse and orthogonal matrix representations directly, without the need for iterative manifold optimization.

\subsubsection{Sparsity Pattern Selection}
For a given sparsity level $s$, which is the number of nonzero elements in a given matrix, we further define the decomposition by considering sparsity patterns, i.e., combinations of nonzero entries that yield column-orthogonal configurations representing points on the complex Grassmann manifold. 
To ensure both sparsity and column orthogonality in a straightforward manner, the following constraints are imposed:
\begin{itemize}
    \item $M\le s\le T$ (sparsity constraint) and
    \item the nonzero entries in different columns are mutually disjoint (column-orthogonal constraint).
\end{itemize}
That is, no row index is shared between two different columns. Under this condition, the columns have disjoint supports, so their inner products automatically vanish. As a result, any such matrix is sparse and {column-orthogonal, i.e., it lies on the complex Grassmann manifold without further adjustment}. Moreover, under this constraint, the number of admissible sparsity patterns can be counted easily, one first chooses $s$ rows out of $T$, and then partitions them into $M$ nonempty disjoint subsets corresponding to the column supports. Thus, for a given $T,M$ and sparsity $s$, its total number of sparsity patterns $n(T,M,s)$ is given by
\begin{align}
    n(T,M,s)=\frac{\binom{T}{s}}{M!}\sum_{k=0}^{M}(-1)^k\binom{M}{k}(M-k)^s
    \label{eq:n(T,M,s)}.
\end{align}

\subsubsection{Embedding Complex Variables}
Next, we select matrices from given $n(T,M,s)$ sparsity patterns. Then, we embed the complex numbers into arbitrary complex parts $\ast$ and perform optimization. At this point, let ${p}$ denote the set of complex parameters for a given matrix $\mathbf{W}$. ${p}$ can be defined as the set of parameters for the amplitude and phase of $s$ complex numbers. 
For example, $p$ can be defined as $p=\{\alpha_1,\theta_1,\cdots\}$, where $\alpha_i,\theta_i$ denote the amplitude and phase of the $i$-th complex elements.
For a given codebook size $|\mathcal{W}|$, we define the set of these parameters as $\mathcal{P}=\{{p}_1,\cdots,{p}_{|\mathcal{W}|}\}$ and perform optimization with respect to $\mathcal{P}$. Here, each parameter can be freely set to both discrete and continuous values  {depending on the application}, as long as the constraints are satisfied. Such sparsity directly reduces the number of nonzero entries in the precoder matrix, thereby lowering implementation complexity and power consumption.

\subsubsection{MCD Maximization}
We optimize the codebook based on the parameters defined above.
{For a codebook} $\mathcal{W} = \{\W_1, \cdots, \W_{|\mathcal{W}|}\}$ {of size} $|\mathcal{W}|$ {with parameter set} $\mathcal{P}=\left\{p_1,\cdots,p_{|\mathcal{W}|} \right\}$, {the optimization objective for maximizing the MCD in (\ref{eq:obj_MCD}) is defined as}
\begin{align}
        \underset{\mathcal{P}}{\operatorname{minimize}}~~&\log \underset{1\leq i<j\leq |\mathcal{W}|}{\sum}\exp\left(-\frac{\|\W_i\W_i^\e-\W_j\W_j^\e\|_\F}{\epsilon} \right)
    \label{eq:obj_1}\\
    \text{s.t.}~~&\mathcal{P} = 
    \left\{p_1,\cdots,p_{|\mathcal{W}|}\right\}, \notag
\end{align}
where $\epsilon$ is a smoothing constant. Each parameter set $p_i \in\mathcal{P}$ is optimized independently under the constraints defined above.

\subsection{Design Method for $T=2M$}
We focus on designing precoding codebooks based on maximizing the MCD under the constraint  $T=2M$. 
This configuration is practically relevant because many uplink precoding scenarios employ twice as many transmit antennas as data streams.
It also provides an analytically symmetric structure that allows explicit evaluation of the chordal distance.
We introduce a method to reduce the search space and simplify the objective function maximizing the MCD in this case.

Here, we assume that all elements of the matrix have equal amplitude and that each matrix is $2M$-sparse, i.e., $s=2M$. Under this constraint, each matrix has exactly two nonzero elements in each of its $M$ columns.
For a given matrix $\W$, we define its sparsity pattern $\pi_i$ as
\begin{align}
 &\pi_i=\{(a_{i,1},b_{i,1}),\cdots,(a_{i,M},b_{i,M})\},
 \label{eq:pi}\\
 \text{s.t}&~\{a_{i,m},b_{i,m}\}~\cap~ \{a_{i,m'},b_{i,m'}\}=\varnothing,~~1\le m<m'\le M \notag,
\end{align}
where each $(a_{i,m},b_{i,m})$ denotes the pair of row indices of the nonzero entries in the $m$-th column vector and $a_{i,m},b_{i,m}\in\{1,\cdots,2M\}$.
Moreover, the row-index pairs collectively cover the entire index set $\{1,\cdots, 2M\}$. That is, 
\begin{align}
    \bigcup_{m=1}^{M}\{a_{i,m},b_{i,m}\}=\{1,\cdots,2M\}
    \label{eq:bigscup_aimbim}.
\end{align}
These conditions imply that the row-index pairs form a partition of $\{1,\cdots,2M\}$ into $M$ disjoint unordered pairs. Hence, each sparsity pattern corresponds to a perfect matching of the $2M$ row indices, which guarantees column orthogonality.
{For a given $T=2M$, we consider a set of sparsity patterns $\pi$ and require that the patterns do not intersect at any column index. Under this constraint, the number of admissible sparsity patterns is $2M-1$.}
That is, for every set $\pi_1,\cdots,\pi_{2M-1}$ of sparsity patterns $\pi_{i}=\{(a_{i,m},b_{i,m})\}^{M}_{m=1}$, we set a constraint as
\begin{align}
    &(a_{i,m},b_{i,m})\neq (a_{j,m},b_{j,m}), \\
    \text{s.t.}&~~1\le i<j\le 2M-1,~m\in\{1,\cdots,M\} \notag.
\end{align}
Consequently, for any two distinct sparsity patterns $\pi_i$ and $\pi_j$, the column-wise index pairs are not identical. Furthermore, since each sparsity pattern forms a perfect matching of $\{1,\cdots,2M\}$, each row index appears exactly once in every pattern. Thus, for a fixed column $m$ of $\pi_i$ with indices $\{a_{i,m}, b_{i,m}\}$, the two indices $a_{i,m}$ and $b_{i,m}$ must belong to two different column pairs of $\pi_j$, implying that the corresponding column vector $\w_{i,m}$ overlaps with exactly two columns of $\W_j$ in one element each.

Then, let $\pi_{i}=\{(a_{i,m},b_{i,m})\}^{M}_{m=1}$ and $\pi_j=\{(a_{j,m},b_{j,m})\}^{M}_{m=1}$ be the corresponding matrices $\W_i=[\w_{i,1},\cdots,\w_{i,M}]$ and $\W_j=[\w_{j,1},\cdots,\w_{j,M}]$. From the Schubert decomposition, each column vector is denoted as $\w_{i,m}=(\a_{a_{i,m}}+e^{\j\theta_{i,m}}\a_{b_{i,m}})/\sqrt{2}$ and $\w_{j,m}=(\a_{a_{j,m}}+e^{\j\theta_{j,m}}\a_{b_{j,m}})/\sqrt{2}$ 
where $\a_i\in\C^{2M}$ denotes the column vector, whose $i$-th element is $1$ and the others are $0$, i.e., the standard basis vector of length $2M$. The inner product of $\W_i$ and $\W_j$ is denoted as
\begin{align}
\|\W_i^\e\W_j\|_\F^2~=\sum_{m=1}^{M}\sum_{m'=1}^M|\langle\w_{i,m},\w_{j,m'}\rangle|^2.
\label{eq:inner_product_1}
\end{align}
Here, since each sparsity pattern is a perfect matching of the same index set, the fixed column $\w_{i,m}$ intersects exactly two columns of $\W_j$ by one element. That is, when the intersection is one element for any $m$ and $m'$, the inner product of each column vector $\w_{i,m}$ and $\w_{j,m'}$ is denoted as
\begin{align}
    \langle\w_{i,m},\w_{j,m'}\rangle&=\frac{1}{\sqrt{2}}e^{-\j\theta_{i,m}}\cdot\frac{1}{\sqrt{2}}e^{\j\theta_{j,m'}} \notag \\
    &=\frac{1}{2}e^{-\j(\theta_{i,m}-\theta_{j,m'})},
\end{align}
\begin{align}
\Leftrightarrow~|\langle\w_{i,m},\w_{j,m'}\rangle|^2&=\left|\frac{1}{2}e^{-\j(\theta_{i,m}-\theta_{j,m'})}\right|^2=\frac{1}{4}.
\label{eq:inner_product_column_1}
\end{align}
{When there is no intersection, the inner product is $0$.} Hence, we obtain
\begin{align}
\sum_{m'=1}^M|\langle\w_{i,m},\w_{j,m'}\rangle|^2=2\cdot\frac{1}{4}=\frac{1}{2}.
\label{eq:inner_product_column_2}
\end{align}
\eqref{eq:inner_product_1} is simplified with the results of~\eqref{eq:inner_product_column_1} and~\eqref{eq:inner_product_column_2} as
\begin{align}
    \|\W_i^\e\W_j\|_\F^2~ &=\sum_{m=1}^{M}\sum_{m'=1}^M|\langle\w_{i,m},\w_{j,m'}\rangle|^2 \notag \\
    &=\sum_{m=1}^{M}\frac{1}{2}=\frac{M}{2}.
    \label{eq:inner_product_2}
\end{align}
From~\eqref{eq:inner_product_2}, the chordal distances between different sparsity patterns are constant, i.e., the chordal distance between $\W_i$ and $\W_j$, $d_{\mathrm{c}}(\W_i,\W_j)$ is derived as
\begin{align}
    d_{\mathrm{c}}(\W_i,\W_j)=\sqrt{M-\|\W_i^\e\W_j\|_\F^2}=\sqrt{\frac{M}{2}},
    \label{eq:dch_diff}
\end{align}
that is, the chordal distance between different sparsity patterns does not depend on the phases of their matrices.
This means that all codewords constructed from distinct sparsity patterns are mutually equidistant on the Grassmann manifold, forming a quasi-uniform codebook structure without explicit manifold optimization.

Next, we consider two matrices $\W_i=[\w_{i,1},\cdots,\w_{i,M}]$ and $\W_k=[\w_{k,1},\cdots,\w_{k,M}]$ that have the same sparsity pattern $\pi_i=\{(a_{i,m},b_{i,m})\}^{M}_{m=1}$. Their column vectors are denoted as $\w_{i,m}=(\a_{a_{i,m}}+e^{\j\theta_{i,m}}\a_{b_{i,m}})/\sqrt{2}$ and $\w_{k,m}=(\a_{a_{k,m}}+e^{\j\theta_{k,m}}\a_{b_{k,m}})/\sqrt{2}$.
{Here, since $\W_i$ and $\W_k$ share the same sparsity pattern $\pi_i$, the inner product between $\w_{i,m}$ and $\w_{k,m'}$ is denoted as}
\begin{align}
    &\langle\w_{i,m},\w_{k,m'}\rangle =
    \begin{dcases}
        \frac{1}{2}\left(1+e^{-\j(\theta_{i,m}-\theta_{k,m'})}\right)~&(m=m')\\
        0&(m\neq m')
    \end{dcases}.
\end{align}
Hence, we obtain $|\langle\w_{i,m},\w_{k,m'}\rangle|^2 =\left( 1+\cos({\theta_{i,m}-\theta_{k,m'}}) \right)/2$ for $m=m'$ and otherwise it is $0$.
From~\eqref{eq:inner_product_1}, the inner product of $\W_i$ and $\W_k$ is simplified as
\begin{align}
    \|\W_i^\e\W_k\|_\F^2 &=\sum_{m=1}^M\sum_{m'=1}^M|\langle\w_{i,m},\w_{k,m'}\rangle|^2 \notag \\
    &= \sum_{m=1}^M|\langle\w_{i,m},\w_{k,m}\rangle|^2 \notag \\
    &= \sum_{m=1}^M \frac{1}{2}\left( 1+\cos({\theta_{i,m}-\theta_{k,m}}) \right) \notag \\
    &= \frac{M}{2}+\frac{1}{2}\sum_{m=1}^M\cos({\theta_{i,m}-\theta_{k,m}})
    \label{eq:inner_product_3}.
\end{align}
Thus, the chordal distance between $\W_i$ and $\W_k$, $ d_{\mathrm{c}}(\W_i,\W_k)$ is denoted as
\begin{align}
    d_{\mathrm{c}}(\W_i,\W_k)&=\sqrt{M-\|\W_i^\e\W_k\|_\F^2} \notag \\
     & = \sqrt{\frac{M}{2}-\frac{1}{2}\sum_{m=1}^M\cos({\theta_{i,m}-\theta_{k,m}})} \notag \\
     &= \sqrt{\sum_{m=1}^M\sin^2{\frac{{\theta_{i,m}-\theta_{k,m}}}{2}}}
     \label{eq:dch_same},
\end{align}
which depends only on the differences in the phase parameters. This confirms that the chordal distance between two matrices with the same sparsity pattern is entirely determined by their relative phase differences. This important result significantly simplifies codebook optimization since all distances between patterns are fixed, and only the phase difference within a pattern affects performance.

{Using the properties}~\eqref{eq:dch_diff} {and}~\eqref{eq:dch_same}, {the MCD-maximization objective for designing a precoding codebook of size $|\mathcal{W}|$, originally defined in~\eqref{eq:obj_1}, can be significantly simplified to a function that depends only on phase differences as}
\begin{align}
    \underset{\theta_{i,m},~\theta_{j,m}}{\operatorname{maximize}}~\underset{1\leq i<j\leq L}{\min}\sum_{m=1}^M\sin^2{\frac{{\theta_{i,m}-\theta_{j,m}}}{2}}
    \label{eq:obj_dch_new},
\end{align}
where $L$ denotes the number of phase instances assigned to each sparsity pattern, which is defined as 
\begin{align}
L=\left\lceil\frac{|\mathcal{W}|}{2M-1}  \right\rceil
\label{eq:L},
\end{align}
where $\lceil x\rceil$ denotes the ceiling function that rounds $x$ up to the nearest integer, and $2M-1$ is the number of distinct sparsity patterns constructed above. Consequently, the codebook size $|\mathcal{W}|$ satisfies $|\mathcal{W}|\le L(2M-1)$. If $|\mathcal{W}|$ is not an integer multiple of $2M-1$, only a subset of sparsity patterns contains $L$ instances, while the remaining patterns contain $L-1$ instances. This uneven assignment does not affect the MCD characterization derived in~\eqref{eq:dch_diff} and~\eqref{eq:dch_same}, as described above.

Compared to the conventional MCD-Manopt approach using (\ref{eq:obj_MCD_2}), we demonstrate that the number of real variables and the associated search space are significantly reduced. Specifically, for a codebook $\mathcal{W} \subset \G(T, M)$ with its codebook size $|\mathcal{W}|$, the number of real variables required in MCD-Manopt is defined as $2|\mathcal{W}|\cdot M(T-M)$. 
In contrast, for the same codebook size under the condition and constraint, the number of real variables is defined as $\lceil |\mathcal{W}|/(2M-1) \rceil\cdot M$ by applying this work. 

\begin{figure}[tb]
    \centering
    \includegraphics[width=\linewidth]{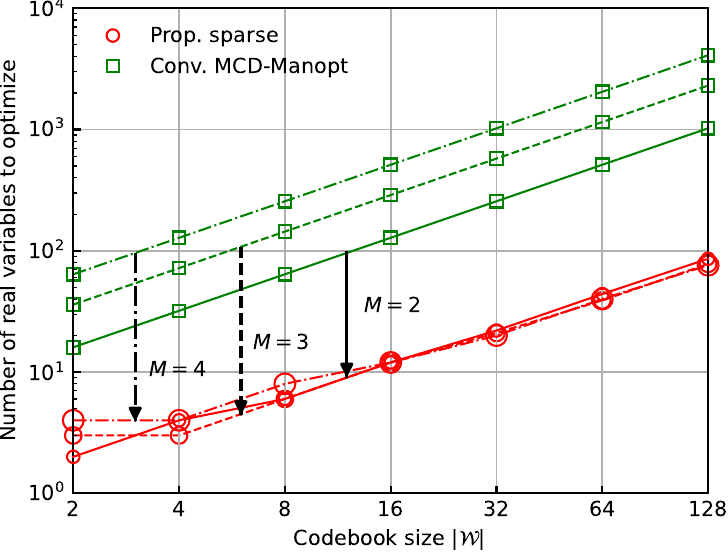}
    \caption{Number of real variables required for optimization when applying the proposed method with the number of transmit antennas $T$ and data streams $M$, where $(T,M)=(4,2),~(6,3),$ and $(8,4)$.}
    \label{fig:variables_comparison}
    \end{figure}

Fig.~\ref{fig:variables_comparison} shows the comparison of the numbers of real variables required for optimization when the proposed simplification method for $2M$-sparse matrices was applied for $(T,M)=(4,2),~(6,3),$ and $(8,4)$. As shown in Fig.~\ref{fig:variables_comparison}, the proposed method required significantly fewer real variables for optimization than MCD-Manopt even for large codebook size in all cases. Furthermore, comparing the results between the methods, the real variables required by MCD-Manopt increased as $M$ increased, whereas those required by the proposed method did not change much as $M$ increased. This result demonstrates that our proposed method is effective even for large matrix sizes that are difficult to optimize.

\section{Numerical Analysis}
\label{sec:analy}
In this section, we analyze {the} numerical performance of the proposed codebook design method. First, we demonstrate that the PAPR varies depending on {precoder sparsity}. Subsequently, we provide a numerical analysis showing how the sparse structure contributes to complexity reduction.
Furthermore, we compare the MCD of the codebook designed in Section~\ref{sec:prop}-B to verify the advantage of the proposed design method. 

\subsection{PAPR Analysis}
We analyze how precoder sparsity affects the PAPR.
To isolate the sparsity effects from codebook optimization, we do not perform further codebook tuning in this section.
We consider a precoder whose nonzero entries have equal amplitude and where each row contains exactly $\ell$ nonzero elements with $\ell\in\{1,\cdots,M\}$ for simplicity. Note that $\ell$ denotes the number of nonzero elements per row, hence the total number of nonzero elements, that is the matrix sparsity level is defined as $s=\ell T$ in this case. Since each element has the same amplitude and each row contains an equal number of nonzero elements, the precoded signals for the $t$-th transmit antenna and $k$-th subcarrier in the frequency-domain $z_t[k]$ are represented as
\begin{align}
    z_t[k]=\sqrt{\frac{M}{\ell T}}\sum_{i=1}^{\ell}e^{\j\theta^{(t)}_i}\cdot s_i[k],
    \label{eq:transmit_PAPR}
\end{align}
where $s_i[k]$ is the data symbol on the $k$-th subcarrier with the index~$i \in \left\{ 1, \cdots, \ell \right\}$ representing the $i$-th non-zero stream among all $M$ streams.
In the OFDM case, the data symbols are drawn from PSK or QAM constellations, whereas in DFT-s-OFDM, they are DFT-spread symbols obtained by applying DFT precoding to a PSK or QAM symbol vector.

\begin{figure*}[tb]
  \centering
  \subfloat[$\ell=1$]{\includegraphics[width=0.24\linewidth]{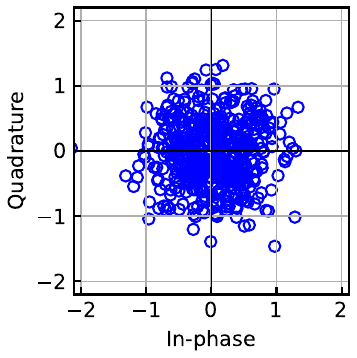}}\hfill
  \subfloat[$\ell=2$]{\includegraphics[width=0.24\linewidth]{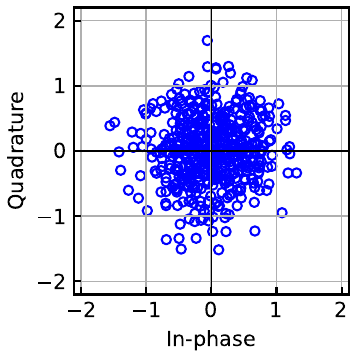}}\hfill
  \subfloat[$\ell=3$]{\includegraphics[width=0.24\linewidth]{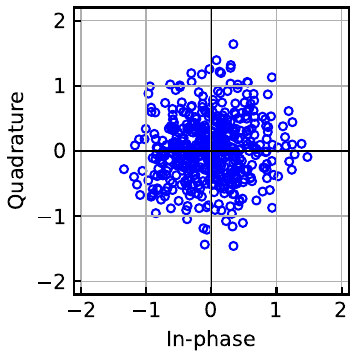}}\hfill
  \subfloat[$\ell=4$]{\includegraphics[width=0.24\linewidth]{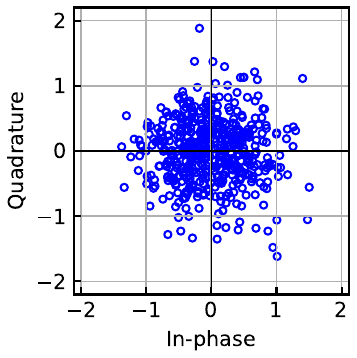}}\hfill
  \caption{Distribution of the time-domain Nyquist rate samples of OFDM
  with $512$ subcarriers and 4-QAM modulation, where $\ell$ denotes the number of nonzero elements in each row of the precoder}
  \label{fig:constellation_OFDM}
\end{figure*}

\begin{figure*}[tb]
  \centering
  \subfloat[$\ell=1$]{\includegraphics[width=0.24\linewidth]{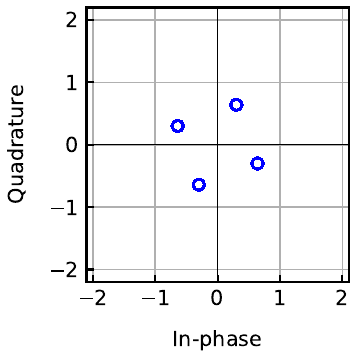}}\hfill
  \subfloat[$\ell=2$]{\includegraphics[width=0.24\linewidth]{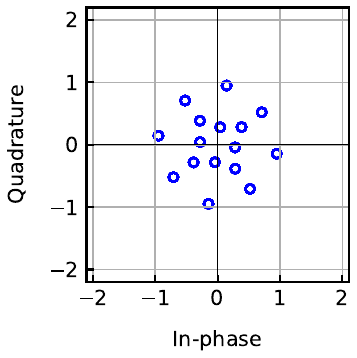}}\hfill
  \subfloat[$\ell=3$]{\includegraphics[width=0.24\linewidth]{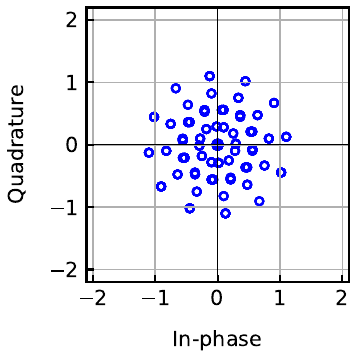}}\hfill
  \subfloat[$\ell=4$]{\includegraphics[width=0.24\linewidth]{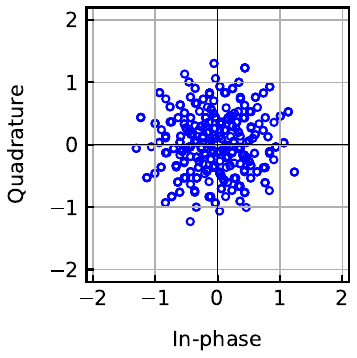}}\hfill
  \caption{Distribution of the time-domain Nyquist rate samples of DFT-s-OFDM with $512$ subcarriers and 4-QAM modulation when varying $\ell$.}
  \label{fig:constellation_DFT}
\end{figure*}

We then perform an inverse fast Fourier transform (IFFT) on~\eqref{eq:transmit_PAPR} and plot the constellation of the resulting time-domain symbols to estimate the instantaneous power of each sample. 
For the case of $(T,M)=(8,4)$, the constellation plots of the time-domain symbols for OFDM and DFT-s-OFDM, as the number of nonzero elements in each row of the precoder $\ell$ varies through $1$ to $4$, are shown in Fig.~\ref{fig:constellation_OFDM} and~\ref{fig:constellation_DFT} respectively. The total number of subcarriers was $512$, and $4$-QAM modulation was used. 
The phase variables $\theta$s were independently drawn from $
[-\pi, \pi]$, yielding $\boldsymbol\theta\triangleq \left[\theta_0^{(t)},\theta_1^{(t)},\theta_2^{(t)},\theta_3^{(t)}\right] = [1.91,-2.21,-1.71,0.636]~\text{rad}$
in this case. For $\ell\in\{1,2,3,4\}$, the first $\ell$ elements were used for mapping.
We note that the oversampling was not applied in Figs.~\ref{fig:constellation_OFDM} and~\ref{fig:constellation_DFT} for the sake of better visualization of the results. 
The trajectories of the time-domain symbols after oversampling in OFDM and DFT-s-OFDM were extremely complex, making it impossible to obtain the necessary information for PAPR analysis. 
Thus, we adopted the time-domain samples obtained when sampling at the Nyquist rate.
Fig.~\ref{fig:constellation_OFDM} shows the constellation distribution for the OFDM case. 
Due to the central limit theorem, the time domain samples of the OFDM signal follow complex Gaussian distribution~\cite{ochiai2001distribution} regardless of the precoder~$\ell$.
Consequently, the PAPR remained nearly constant and was independent of the sparsity.
Fig.~\ref{fig:constellation_DFT} shows the constellation distribution for DFT-s-OFDM. 
The operation of the precoder is essentially equivalent to superimposing single-carrier signals. Therefore, a finite constellation formed by the superposition of $\ell$ $4$-QAM symbols is generated as a result.
As shown in~\eqref{eq:transmit_PAPR}, when $\ell=1$, the constellation was equivalent to phase rotation of the input symbols and thus did not affect PAPR. As $\ell$ increased, the constellation distribution became denser and more irregular, leading to higher peaks and larger variance, and consequently, an increase in PAPR. In the limit, as the constellation becomes dense enough, it theoretically approaches a complex Gaussian distribution. 
Although the results depend on the input symbols and the specific precoder, on average, similar tendencies are observed as shown in Fig.~\ref{fig:constellation_OFDM} and Fig.~\ref{fig:constellation_DFT}.
In principle, the continuous-time signal should be reconstructed by oversampling, and the PAPR should be evaluated accordingly. However, to highlight that the dense precoder increases the PAPR through the superposition of modulated signal points in DFT-s-OFDM, we restrict our analysis here to samples taken at the Nyquist rate. In what follows, the PAPR is discussed based on the continuous-time signal reconstructed by oversampling.
\begin{figure}[tb]
    \centering
    \includegraphics[width=\linewidth]{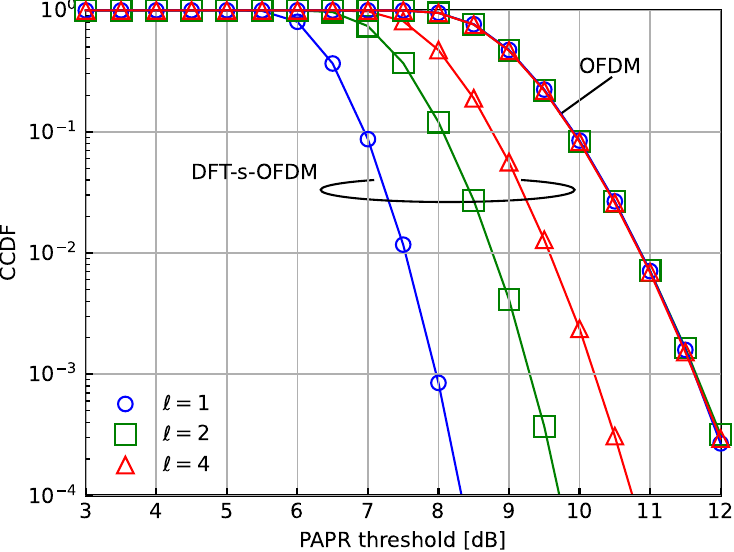}
    \caption{PAPR dependency in OFDM and DFT-s-OFDM with $512$ subcarriers and 4-QAM modulation when varying the number of nonzero elements in each row of the precoder $\ell=1$, $2$, $3$ and $4$.}
    \label{fig:PAPR_independency}
\end{figure}

To statistically characterize the PAPR behavior, we employed the complementary cumulative distribution function (CCDF), which represents the probability that the PAPR of a transmit signal exceeds a given threshold that is defined as 
\begin{align}
    \text{CCDF}(\text{PAPR}_0)=\text{Pr}(\text{PAPR}>\text{PAPR}_0).
    \label{eq:CCDF}
\end{align}
Fig.~\ref{fig:PAPR_independency} shows {the} CCDF of PAPR when using $512$ subcarriers and $4$-QAM modulation. We performed $8$ times oversampling for accurate PAPR estimation in CCDF evaluation. As shown in Fig.~\ref{fig:PAPR_independency}, for DFT-s-OFDM, the PAPR decreased as the precoder became sparser, whereas for OFDM, the PAPR remained almost independent of sparsity.
This confirms the validity of the view that in OFDM, the constellation does not depend on sparsity and does not affect PAPR, whereas in DFT-s-OFDM, the constellation becomes more structured as it becomes sparse, leading to a decrease in PAPR. Owing to the constraints introduced in Section~\ref{sec:prop}, the proposed codebooks are required to satisfy $\ell=1$ to ensure the column-orthogonality condition. As discussed above, the proposed codebook achieves the minimum possible PAPR. 

\subsection{Complexity Analysis}
Next, in this work, we analyze the computational complexity of the proposed sparse precoders in comparison with conventional dense precoders in terms of the number of complex multiplications.
For simplicity, we assume here that $\ell=1$ in the definition of~\eqref{eq:transmit_PAPR}.
The complexity evaluation is divided into three parts: precoder selection, precoder multiplication and space complexity.
For precoder selection, we select the optimal precoder index using the achievable rate given in~\eqref{eq:capacity_i},
which includes the Gram matrix computation of~$\left( \mathbf{HW} \right)^\text{H} \left( \mathbf{HW} \right)$.
For dense matrices $\W \in \C^{T\times M}$ and $\H \in \C^{N\times T}$, the computation of~$\H\W$ requires $NTM$ complex multiplications.
Thus, the Gram matrix computation requires $NTM+NM^2$ in total.
On the other hand,
since each row of the proposed sparse precoder contains only one non-zero entry,
the number of complex multiplications in the computation of~$\mathbf{HW}$ is reduced to~$NT$.
Thus, the total complexity of the Gram matrix computation~$\left( \mathbf{HW} \right)^\text{H} \left( \mathbf{HW} \right)$
with sparse precoders is given by~$NT+NM^2$.

For precoder multiplication,
the transmitter applies the selected precoder~$\mathbf{W} \in \mathbb{C}^{T \times M}$ to the modulated signal vector~$\mathbf{s} \in \mathbb{C}^{M \times 1}$ before transmission.
This operation requires $MT$ complex multiplications in general,
but it can be reduced to~$T$ with our proposed sparse precoders.

Furthermore, the proposed sparse codebooks also {reduce} the space complexity. Data structures for sparse matrices are summarized in~\cite{bell2008efficient}. Owing to the constraints defined in Section~\ref{sec:prop}, each row of the proposed constellation matrix stores only one nonzero element, and thus the well-known ELLPACK format~\cite{kincaid1989ITPACKV} can be used. For a general dense matrix with the same matrix dimensions and codebook size, the space complexity is $\mathcal{O}(|\mathcal{W}|TM)$. In contrast, for the proposed sparse constellation, employing the ELLPACK format reduces the space complexity to $\mathcal{O}(|\mathcal{W}|T)$, because only the positions of the nonzero elements and their complex values need to be stored. Hence, similar to the time complexity, the space complexity is also reduced to linear order with respect to $M$.

Thus, with sparsity the computational cost scales linearly rather than quadratically with antenna dimension, providing efficiency in wideband MIMO systems.
\begin{table}[tb]
    \centering
    \caption{Computational complexity comparisons}
    \label{tab:complexity}
    \begin{tabular}{|c|c|c|} \hline
          & Conv. dense & Prop. sparse ($\ell=1$) \\ \hline
        Index selection & $\mathcal{O}((TM+M^2)N)$ & $\mathcal{O}((T+M^2)N)$ \\ \hline 
        Precoder multiplication & $\mathcal{O}(TM)$ & $\mathcal{O}(T)$ \\ \hline
        Space complexity & $\mathcal{O}(TM|\mathcal{W}|)$ & $\mathcal{O}(T|\mathcal{W}|)$ \\ \hline
    \end{tabular}
\end{table}

\subsection{Comparisons of the MCD}
\begin{figure}[!tb]
  \centering
  \subfloat[$(T,M)=(4,2)$.]{%
    \includegraphics[width=\linewidth]{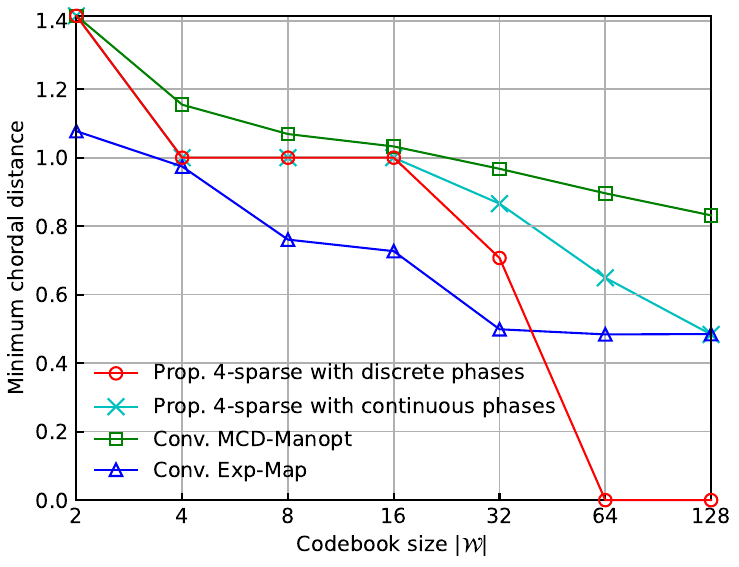}%
    \label{fig:mcd_comparison_M=2_T=4}%
  }
  \vspace{1mm}
  \subfloat[$(T,M)=(6,3)$.]{%
    \includegraphics[width=\linewidth]{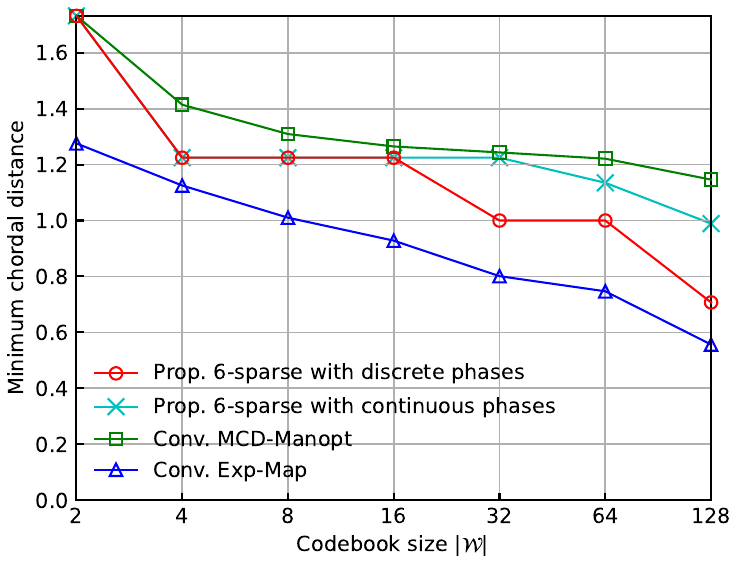}%
    \label{fig:mcd_comparison_M=3_T=6}%
  }
  \vspace{1mm}
  \subfloat[$(T,M)=(6,2)$.]{%
    \includegraphics[width=\linewidth]{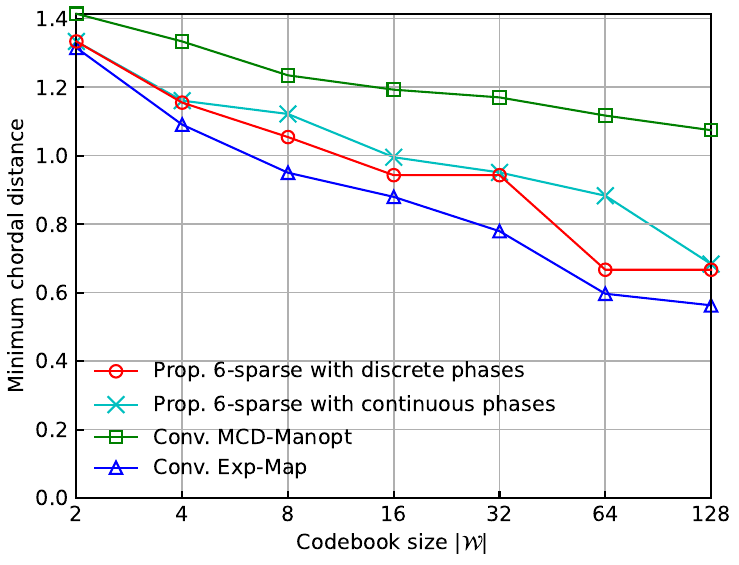}%
    \label{fig:mcd_comparison_M=2_T=6}%
  }
  \caption{Minimum chordal distance comparisons with the number of transmit antennas $T$ and data streams $M$, where $(T,M)=(4,2)$, $(6,3)$, and $(6,2)$.}
  \label{fig:MCD_comparison}
\end{figure}

We compare the proposed sparse codebook and conventional codebooks, MCD-Manopt and Exp-Map in terms of the MCD. Fig.~\ref{fig:MCD_comparison}~\subref{fig:mcd_comparison_M=2_T=4} shows the result for the case of $(T,M)=(4,2)$. The sparse codebooks were designed according to the method described in Section~\ref{sec:prop}-B. For the discrete case, phase variables were quantized as $\theta \in \left\{-\pi/2,~0,~\pi/2,~\pi \right\}$. For any codebook size $|\mathcal{W}|$, the proposed codebooks outperformed Exp-Map and achieved performance comparable to MCD-Manopt. Fig.~\ref{fig:MCD_comparison}~\subref{fig:mcd_comparison_M=3_T=6} presents result for $(T,M)=(6,3)$, where the proposed codebook was constructed in the same procedure as in Fig.~\ref{fig:MCD_comparison}~\subref{fig:mcd_comparison_M=2_T=4}. The proposed method outperformed Exp-Map and approached the performance of MCD-Manopt. These results indicate that the proposed method is effective even when the dimensions of the matrices $T$ and $M$ increase. Finally, Fig.~\ref{fig:MCD_comparison}~\subref{fig:mcd_comparison_M=2_T=6} shows the result when $(T,M)=(6,2)$. In this case, the simplification method introduced in Section~\ref{sec:prop}-B cannot be applied. Nevertheless, sparse codebooks still achieved a larger MCD than Exp-Map for all values of $|\mathcal{W}|$. 

Across all tested dimensions, the proposed sparse design consistently delivered MCD values approaching the theoretical optimal results obtained by MCD-Manopt,
while maintaining a structural simplicity.

\section{Performance Comparisons}
\label{sec:res}
In this section, we investigate the performance of the codebooks designed by the proposed method. We evaluated the achievable rate under uncorrelated Rayleigh fading channels, the effective channel gain under Rician fading channels, and the PAPR in OFDM and DFT-s-OFDM. 
In the achievable rate comparisons and the effective channel gain comparison, the number of receive antennas was set to $N=32$ assuming uplink scenario.

\subsection{Comparisons of Achievable Rate}
We first compare the proposed sparse codebook with MCD-Manopt and the codebook currently employed in 5G NR systems defined in the Table 6.3.1.5-5 and 6.3.1.5-12 in~\cite{3gpp_ts_38211_v1820}. In terms of the achievable rate, we calculated the expectation of~\eqref{eq:capacity_i}, and each index $i$ was selected according to the method in~\eqref{eq:selection}. We constructed the codebook with $2$ and $4$-sparse matrices for $(T,M)=(4,2)$, and $8$-sparse matrices for $(T,M)=(8,4)$ based on the criteria noted in Section~\ref{sec:prop}-B.
The codebook size $|\mathcal{W}|$ was set to $22$ for $(T,M)=(4,2)$ and $24$ for $(T,M)=(8,4)$.
The amplitude of each element in the matrix was set to a discrete values of $\pm1$ and $\pm \j$, consistent with the 5G NR specification in~\cite{3gpp_ts_38211_v1820}.

\begin{figure}[tb]
    \centering
    \includegraphics[width=\linewidth]{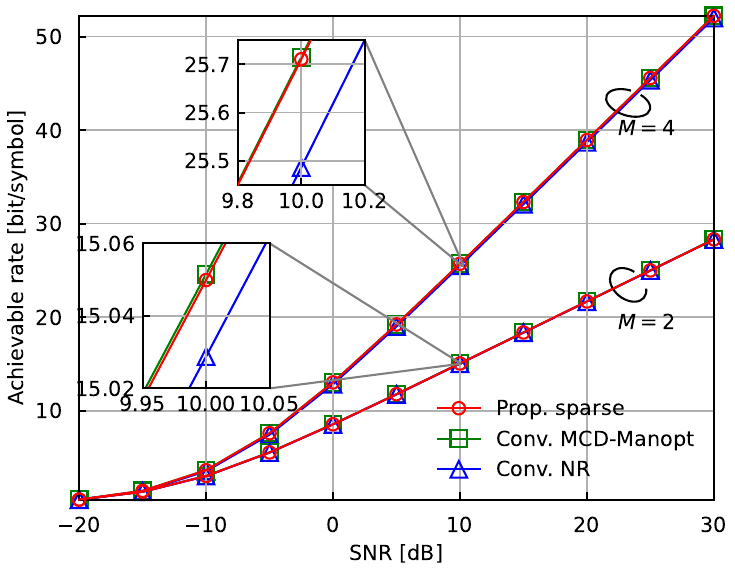}
    \caption{Achievable rate under uncorrelated Rayleigh fading channels with $(T,M)=(4,2)$ and $(8,4)$.}
    \label{fig:achievable_rate_32}
\end{figure}

Fig.~\ref{fig:achievable_rate_32} shows the results of the achievable rate of three codebooks with the same codebook size under uncorrelated Rayleigh fading channels. 
As illustrated in Fig.~\ref{fig:achievable_rate_32},
the proposed sparse codebook achieved almost identical achievable rates to the optimal MCD-Manopt codebook across the entire SNR range. The proposed codebook exhibited no measurable SNR loss, requiring less than $0.01$ $\text{dB}$ of  additional SNR compared with MCD-Manopt. In contrast, it achieved an SNR gain of approximately $0.025$ $\text{dB}$ over the 5G NR codebook.
These results indicate that the proposed sparse method maintains the Grassmannian optimality of MCD-Manopt while providing a consistent SNR advantage over the NR codebook under uncorrelated Rayleigh fading channels.

As illustrated in Fig.~\ref{fig:achievable_rate_32}, the SNR gain of the proposed sparse codebook became more pronounced as the system dimension increased.
The proposed design achieved an SNR gain of about $0.17$ $\text{dB}$ relative to the 5G NR codebook, while maintaining a negligible gap with MCD-Manopt.
This demonstrates that the performance advantage of the proposed method scales with the number of antennas and data streams, confirming its effectiveness for large-scale MIMO precoding.

As we noted in Section~\ref{sec:analy}-B, the sparse structure also reduces the computational complexity of evaluating~\eqref{eq:capacity_i} and performing the index selection in~\eqref{eq:selection}, since it contains several zero elements compared to dense codebooks. Notably, even under this discrete-value restriction, the proposed codebook achieved superior performance, highlighting both its practical relevance and efficiency.

\subsection{Comparison of Effective Channel Gain of Rician Fading}
We compare the effective channel gain under Rician fading channels to verify the properties exhibited by the proposed codebook across various channel models. The Rician channel is defined in Section~\ref{sec:sys}. We compared three cases of $K$-factor, $K=0$ (NLoS only), $K=1$ (NLoS and LoS equal in proportion), and $K=\infty$ (LoS only).
For the precoder matrix $\W_i$, the effective channel gain of $\H$ is given by $\|\H\W_i\|_\F^2$. Each index $i$ was selected from the codebook to maximize the effective channel gain for the given $\H$. All channels are normalized such that $\E[\|\H\|_\F^2]=1$ for ease of evaluation.

\begin{figure}[tb]
    \centering
    \includegraphics[width=\linewidth]{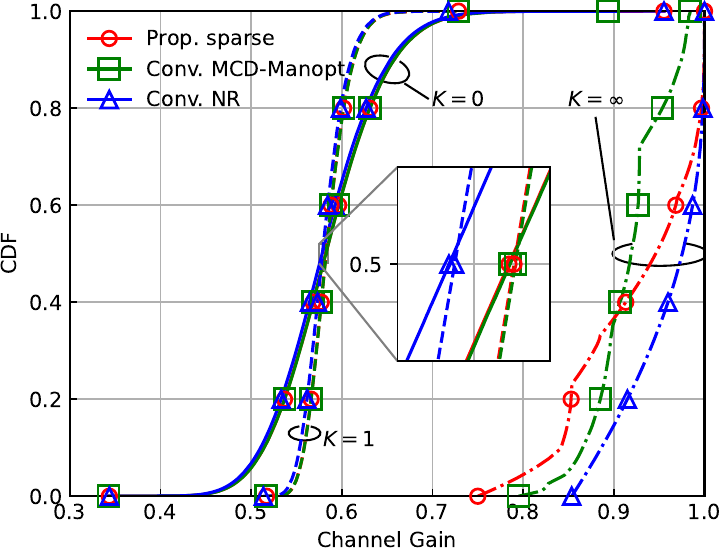}
    \caption{Effective channel gain in Rician channels with $(T,M,N)=(4,2)$, where the Rician factor was $K=0, 1,$ and $\infty$.}
    \label{fig:Rician_CDF}
\end{figure}

In Fig.~\ref{fig:Rician_CDF}, to statistically characterize the performance of the precoding codebooks, we evaluated the empirical cumulative distribution function (CDF) of the effective channel gain over a large number of independent channel realizations.
The same codebooks as those used in the achievable rate comparison for $(T,M)=(4,2)$, were employed here.
Fig.~\ref{fig:Rician_CDF} shows that for $K = 0$ and $K = 1$, the proposed codebook achieved performance comparable to the theoretically optimal MCD-Manopt and outperformed the NR codebook. This demonstrates that the design criterion of ``maximizing the MCD within the codebook'' is effective when the scattered component dominates, confirming that the proposed codebook exhibits comparable performance.
In contrast, when the LoS component became dominant, this criterion was less effective, resulting in both the proposed sparse codebook and MCD-Manopt being inferior to the NR codebook. Nevertheless, the proposed codebook still outperformed MCD-Manopt under LoS-dominant conditions.
These results suggest that the proposed sparse codebook offers a practical and robust alternative to MCD-Manopt.
It maintains near-optimal performance in uncorrelated Rayleigh and moderately Rician environments (i.e., $K \le 1$)
and continues to perform competitively even when the channel becomes LoS-dominant,
where NR codebook naturally has an advantage.

\subsection{Comparisons of PAPR}
\begin{figure}[tb]
    \centering
    \includegraphics[width=\linewidth]{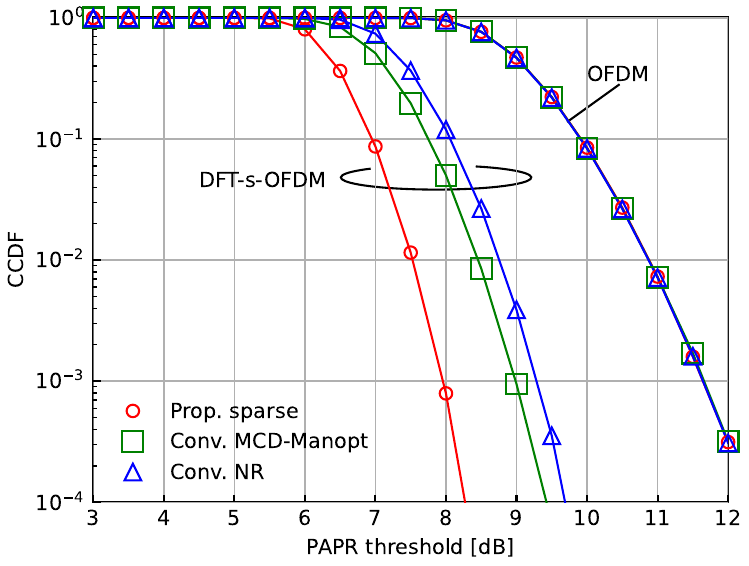}
    \caption{Comparison of PAPR performance in OFDM and DFT-s-OFDM with $(T,M)=(4,2)$, a codebook size of $8$, $52$ PRBs (624 subcarriers), $1024$-point FFT, and $8$ times oversampling.}
    \label{fig:PAPR_average}
\end{figure}

To demonstrate that the proposed codebook is effective in a realistic 5G uplink system, in Fig.~\ref{fig:PAPR_average}, we evaluated PAPR using CCDF. The PAPR was computed for the time-domain signal at each antenna under both OFDM and DFT-s-OFDM, and the average across antennas was reported in Fig.~\ref{fig:PAPR_average}. The simulations assumed $52$ physical resource blocks (PRBs), i.e., $624$ subcarriers with an FFT size of $1024$. To ensure accurate PAPR estimation, time-domain signals were generated with $8$ times oversampling. The codebook size was set to $8$ which has different sparsity from the NR codebook because the PAPR only depends on the sparsity as described in Section~\ref{sec:analy}-A.  

Fig.~\ref{fig:PAPR_average} indicates that the proposed sparse codebook yielded a substantially lower PAPR than both the dense MCD-Manopt codebook and the 5G NR codebook under DFT-s-OFDM transmission. In particular, the proposed codebook achieved a PAPR reduction of more than $1$~dB relative to both dense codebooks at $\text{CCDF}=10^{-2}$.  By contrast, as we noted in Section~\ref{sec:analy}-A, under OFDM transmission, the three codebooks exhibited nearly identical PAPR, which was expected because, as the number of subcarriers increases, the aggregate signal approached a Gaussian distribution by the central limit theorem and the PAPR became dominated by the multi-carrier structure rather than by the precoder.

Overall, the results in Fig.~\ref{fig:PAPR_average},
together with those in Fig.~\ref{fig:achievable_rate_32} and Fig.~\ref{fig:Rician_CDF},
demonstrate that the proposed sparse codebook maintains high achievable rate performance and also provides a notable PAPR advantage in DFT-s-OFDM uplink systems operating under PA constraints.
This reduction in PAPR effectively allows the transmitter to operate with higher average power without increasing distortion, which can directly translate into an achievable rate improvement at the same power budget.
Even apart from the PAPR perspective, these findings confirmed that the proposed sparse design was a practical and power-efficient alternative
to conventional dense codebooks for a wide range of uplink transmission scenarios.
 
\section{Conclusion}
\label{sec:conc}
In this paper, we proposed a sparse Grassmannian codebook design that achieves low complexity and improved power efficiency for limited-feedback MIMO systems. By incorporating Schubert cell decomposition and enforcing sparsity in the precoder structure, the proposed method simplified the objective function and reduces the optimization search space in a specific case. Numerical results showed that the proposed sparse codebooks attained nearly identical achievable rate performance to the theoretically optimal dense MCD-Manopt codebook, while providing substantially lower PAPR in DFT-s-OFDM systems. Moreover, the sparse structure inherently reduces precoder multiplication, index selection, and space complexity, making the design suitable for large-scale and power-constrained transmission scenarios such as IoT and uplink systems. Overall, the proposed sparse codebook design provides an effective framework for practical Grassmannian codebooks, achieving a balanced trade-off among performance, complexity, and power efficiency, and has the potential to replace existing NR codebooks in multi-stream DFT-s-OFDM systems toward 6G.

\appendices
\section{Codebook used in this study}
\label{app:codebook}

The codebooks used in these simulations are given in Table~\ref{tab:NR_TM=42} and~\ref{tab:prop_TM=42}. Although Table~\ref{tab:NR_TM=42} is cited from~\cite{3gpp_ts_38211_v1820}, note that it is normalized to be $\W^\e\W=\I_M$ for each index, which is different from the actual values. The matrices with indices 1-6, i.e., the $2$-sparse matrices coincide in Table~\ref{tab:NR_TM=42} and~\ref{tab:prop_TM=42} because the Schubert cells given in~\eqref{eq:Schubert_decompose} automatically limit the 2-sparse matrices to these six. The other matrices are designed by the method introduced in Section~\ref{sec:prop}-B.
\begin{table}[tb]
    \centering
    \caption{NR Codebook with $(T,M)=(4,2)$.}
    \label{tab:NR_TM=42}
    \begin{tabular}{|c|c|c|c|}\hline
       \textbf{Index}  &  \multicolumn{3}{|c|}{\textbf{Precoder} $\W$} \\ \hline
        1-3 & $\begin{bmatrix}
            1 & 0 \\
            0 & 1 \\
            0 & 0 \\
            0 & 0
        \end{bmatrix}$ & $\begin{bmatrix}
            1 & 0 \\
            0 & 0 \\
            0 & 1 \\
            0 & 0
        \end{bmatrix}$ & $\begin{bmatrix}
            1 & 0 \\
            0 & 0 \\
            0 & 0 \\
            0 & 1
        \end{bmatrix}$ \\ \hline
        
        4-6 & $\begin{bmatrix}
            0 & 0 \\
            1 & 0 \\
            0 & 1 \\
            0 & 0
        \end{bmatrix}$ & $\begin{bmatrix}
            0 & 0 \\
            1 & 0 \\
            0 & 0 \\
            0 & 1
        \end{bmatrix}$ & $\begin{bmatrix}
            0 & 0 \\
            0 & 0 \\
            1 & 0 \\
            0 & 1
        \end{bmatrix}$ \\ \hline
        
        7-9 & $\dfrac{1}{\sqrt{2}}\begin{bmatrix}
            1 & 0 \\
            0 & 1 \\
            1 & 0 \\
            0 & -\j
        \end{bmatrix}$ & $\dfrac{1}{\sqrt{2}}\begin{bmatrix}
            1 & 0 \\
            0 & 1 \\
            1 & 0 \\
            0 & \j
        \end{bmatrix}$ & $\dfrac{1}{\sqrt{2}}\begin{bmatrix}
            1 & 0 \\
            0 & 1 \\
            -\j & 0 \\
            0 & 1
        \end{bmatrix}$ \\ \hline

        10-12 & $\dfrac{1}{\sqrt{2}}\begin{bmatrix}
            1 & 0 \\
            0 & 1 \\
            -\j & 0 \\
            0 & -1
        \end{bmatrix}$ & $\dfrac{1}{\sqrt{2}}\begin{bmatrix}
            1 & 0 \\
            0 & 1 \\
            -1 & 0 \\
            0 & -\j
        \end{bmatrix}$ & $\dfrac{1}{\sqrt{2}}\begin{bmatrix}
            1 & 0 \\
            0 & 1 \\
            -1 & 0 \\
            0 & \j
        \end{bmatrix}$ \\ \hline

        13-15 & $\dfrac{1}{\sqrt{2}}\begin{bmatrix}
            1 & 0 \\
            0 & 1 \\
            \j & 0 \\
            0 & 1
        \end{bmatrix}$ & $\dfrac{1}{\sqrt{2}}\begin{bmatrix}
            1 & 0 \\
            0 & 1 \\
            \j & 0 \\
            0 & -1
        \end{bmatrix}$ & $\dfrac{1}{{2}} \begin{bmatrix}
            1 & 1 \\
            1 & 1 \\
            1 & -1 \\
            1 & -1
        \end{bmatrix}$ \\ \hline

        16-18 & $\dfrac{1}{{2}} \begin{bmatrix}
            1 & 1 \\
            1 & 1 \\
            \j & -\j \\
            \j & -\j
        \end{bmatrix}$ & $\dfrac{1}{{2}} \begin{bmatrix}
            1 & 1 \\
            \j & \j \\
            1 & -1 \\
            \j & -\j
        \end{bmatrix}$ & $\dfrac{1}{{2}} \begin{bmatrix}
            1 & 1 \\
            \j & \j \\
            \j & -\j \\
            -1 & 1
        \end{bmatrix}$ \\ \hline

        19-21 & $\dfrac{1}{{2}} \begin{bmatrix}
            1 & 1 \\
            -1 & -1 \\
            1 & -1 \\
            -1 & 1
        \end{bmatrix}$ & $\dfrac{1}{{2}} \begin{bmatrix}
            1 & 1 \\
            -1 & -1 \\
            \j & -\j \\
            -\j & \j
        \end{bmatrix}$ & $\dfrac{1}{{2}} \begin{bmatrix}
            1 & 1 \\
            -\j & -\j \\
            1 & -1 \\
            -\j & \j
        \end{bmatrix}$ \\ \hline

        22 & $\dfrac{1}{{2}} \begin{bmatrix}
            1 & 1 \\
            -\j & -\j \\
            \j & -\j \\
            1 & -1
        \end{bmatrix}$ &  & \\ \hline
    \end{tabular}
\end{table}

\begin{table}[tb]
    \centering
    \caption{Proposed Codebook with $(T,M)=(4,2)$.}
    \label{tab:prop_TM=42}
    \begin{tabular}{|c|c|c|c|}\hline
       \textbf{Index}  &  \multicolumn{3}{|c|}{\textbf{Precoder} $\W$} \\ \hline
        1-3 & $\begin{bmatrix}
            1 & 0 \\
            0 & 1 \\
            0 & 0 \\
            0 & 0
        \end{bmatrix}$ & $\begin{bmatrix}
            1 & 0 \\
            0 & 0 \\
            0 & 1 \\
            0 & 0
        \end{bmatrix}$ & $\begin{bmatrix}
            1 & 0 \\
            0 & 0 \\
            0 & 0 \\
            0 & 1
        \end{bmatrix}$ \\ \hline
        
        4-6 & $\begin{bmatrix}
            0 & 0 \\
            1 & 0 \\
            0 & 1 \\
            0 & 0
        \end{bmatrix}$ & $\begin{bmatrix}
            0 & 0 \\
            1 & 0 \\
            0 & 0 \\
            0 & 1
        \end{bmatrix}$ & $\begin{bmatrix}
            0 & 0 \\
            0 & 0 \\
            1 & 0 \\
            0 & 1
        \end{bmatrix}$ \\ \hline
        
        7-9 & $\dfrac{1}{\sqrt{2}}\begin{bmatrix}
            1 & 0 \\
            0 & 1 \\
            -\j & 0 \\
            0 & 1
        \end{bmatrix}$ & $\dfrac{1}{\sqrt{2}}\begin{bmatrix}
            1 & 0 \\
            0 & 1 \\
            1 & 0 \\
            0 & \j
        \end{bmatrix}$ & $\dfrac{1}{\sqrt{2}}\begin{bmatrix}
            1 & 0 \\
            0 & 1 \\
            1 & 0 \\
            0 & -\j
        \end{bmatrix}$ \\ \hline

        10-12 & $\dfrac{1}{\sqrt{2}}\begin{bmatrix}
            1 & 0 \\
            0 & 1 \\
            -\j & 0 \\
            0 & -1
        \end{bmatrix}$ & $\dfrac{1}{\sqrt{2}}\begin{bmatrix}
            1 & 0 \\
            0 & 1 \\
            \j & 0 \\
            0 & -1
        \end{bmatrix}$ & $\dfrac{1}{\sqrt{2}}\begin{bmatrix}
            1 & 0 \\
            0 & 1 \\
            -1 & 0 \\
            0 & \j
        \end{bmatrix}$ \\ \hline

        13-15 & $\dfrac{1}{\sqrt{2}}\begin{bmatrix}
            1 & 0 \\
            0 & 1 \\
            \j & 0 \\
            0 & 1
        \end{bmatrix}$ & $\dfrac{1}{\sqrt{2}}\begin{bmatrix}
            1 & 0 \\
            0 & 1 \\
            -1 & 0 \\
            0 & -\j
        \end{bmatrix}$ &  $\dfrac{1}{\sqrt{2}}\begin{bmatrix}
            1 & 0 \\
            -1 & 0 \\
            0 & 1 \\
            0 & -1
        \end{bmatrix}$ \\ \hline

        16-18 & $\dfrac{1}{\sqrt{2}}\begin{bmatrix}
            1 & 0 \\
            \j & 0 \\
            0 & 1 \\
            0 & -\j
        \end{bmatrix}$ &$\dfrac{1}{\sqrt{2}}\begin{bmatrix}
            1 & 0 \\
            -\j & 0 \\
            0 & 1 \\
            0 & -\j
        \end{bmatrix}$ & $\dfrac{1}{\sqrt{2}}\begin{bmatrix}
            1 & 0 \\
            1 & 0 \\
            0 & 1 \\
            0 & \j
        \end{bmatrix}$ \\ \hline

        19-21 & $\dfrac{1}{\sqrt{2}}\begin{bmatrix}
            1 & 0 \\
            0 & 1 \\
            0 & -1 \\
            -1 & 0
        \end{bmatrix}$ & $\dfrac{1}{\sqrt{2}}\begin{bmatrix}
            1 & 0 \\
            0 & 1 \\
            0 & -\j \\
            \j & 0
        \end{bmatrix}$ & $\dfrac{1}{\sqrt{2}}\begin{bmatrix}
            1 & 0 \\
            0 & 1 \\
            0 & -\j \\
            -\j & 0
        \end{bmatrix}$ \\ \hline

        22 & $\dfrac{1}{\sqrt{2}}\begin{bmatrix}
            1 & 0 \\
            0 & 1 \\
            0 & \j \\
            1 & 0
        \end{bmatrix}$ &  & \\ \hline
    \end{tabular}
\end{table}

\footnotesize{
	\bibliographystyle{IEEEtranURLandMonthDiactivated}
	\bibliography{main}

@article{ayach2014spatially,
   author={Ayach, Omar El and Rajagopal, Sridhar and Abu-Surra, Shadi and Pi, Zhouyue and Heath, Robert W.},
  title   = {Spatially sparse precoding in millimeter wave {MIMO} systems},
  journal = {IEEE Transactions on Wireless Communications},
  volume  = {13},
  number  = {3},
  pages   = {1499--1513},
  month   = mar,
  year    = {2014}
}

@article{bendokat2024grassmann,
  title = {A {{Grassmann}} Manifold Handbook: Basic Geometry and Computational Aspects},
  author = {Bendokat, Thomas and Zimmermann, Ralf and Absil, P.},
  year = 2024,
  month = feb,
  journal = {Advances in Computational Mathematics},
  volume = {50},
  number = {1},
  pages = {6},
  doi = {10.1007/s10444-023-10090-8}
}

@article{berardinelli2008ofdma,
   author={Berardinelli, Gilberto and Ruiz de Temino, Luis Angel Maestro and Frattasi, Simone and Rahman, Muhammad Imadur and Mogensen, Preben},
  journal={IEEE Wireless Communications}, 
  title={{OFDMA} vs. {SC-FDMA}: {Performance} comparison in local area {IMT-A} scenarios}, 
  year={2008},
  volume={15},
  number={5},
  pages={64-72},
  doi={10.1109/MWC.2008.4653134}}

@article{berardinelli2009feasibility,
  author  = {G. Berardinelli and Ruiz de Temino, Luis Angel Maestro and S. Frattasi and T. B. S{\o}rensen and P. Mogensen and K. Pajukoski},
  title   = {On the feasibility of precoded single-user {MIMO} for {LTE-A} uplink},
  journal = {Journal of Communications},
  volume  = {4},
  number  = {3},
  pages   = {155--163},
  year    = {2009}
}

@article{conway1996packing,
  title = {Packing Lines, Planes, Etc.: Packings in {{Grassmannian}} Spaces},
  author = {Conway, John H. and Hardin, Ronald H. and Sloane, Neil J. A.},
  year = 1996,
  month = jan,
  journal = {Experimental Mathematics},
  volume = {5},
  number = {2},
  pages = {139--159},
  doi = {10.1080/10586458.1996.10504585}
}

@article{cuevas2023union,
  title = {Union Bound Minimization Approach for Designing {{Grassmannian}} Constellations},
  author = {Cuevas, Diego and {A}lvarez-Vizoso, Javier and Beltr{a}n, Carlos and Santamaria, Ignacio and Tuvcek, Vit and Peters, Gunnar},
  year = 2023,
  month = apr,
  journal = {IEEE Transactions on Communications},
  volume = {71},
  number = {4},
  pages = {1940--1952},
  publisher = {{Institute of Electrical and Electronics Engineers (IEEE)}},
  doi = {10.1109/tcomm.2023.3244965},
}

@article{cuevas2024constellations,
  title = {Constellations on the Sphere With Efficient Encoding-Decoding for Noncoherent Communications},
  author = {Cuevas, Diego and {A}lvarez-Vizoso, Javier and Beltr{a}n, Carlos and Santamaria, Ignacio and Tu{v c}ek, V{i}t and Peters, Gunnar},
  year = 2024,
  month = mar,
  journal = {IEEE Transactions on Wireless Communications},
  volume = {23},
  number = {3},
  pages = {1886--1898},
  doi = {10.1109/TWC.2023.3292935},
}

@article{endo2024boosting,
  title = {Boosting Spectral Efficiency with Data-Carrying Reference Signals on the {{Grassmann}} Manifold},
  author = {Endo, Naoki and Iimori, Hiroki and Pradhan, Chandan and Malomsoky, Szabolcs and Ishikawa, Naoki},
  year = 2024,
  month = aug,
  journal = {IEEE Transactions on Wireless Communications},
  volume = {23},
  number = {8},
  pages = {10137--10149},
  doi = {10.1109/TWC.2024.3369091},
}

@article{gohary2009noncoherent,
  title = {Noncoherent {{MIMO}} Communication: {{Grassmannian}} Constellations and Efficient Detection},
  author = {Gohary, Ramy H. and Davidson, Timothy N.},
  year = 2009,
  month = mar,
  journal = {IEEE Transactions on Information Theory},
  volume = {55},
  number = {3},
  pages = {1176--1205},
  doi = {10.1109/TIT.2008.2011512},
}

@article{hama2023timefrequency,
  title = {Time-Frequency Domain Non-Orthogonal Multiple Access for Power Efficient Communications},
  author = {Hama, Yuto and Ochiai, Hideki},
  year = 2023,
  month = sep,
  journal = {IEEE Transactions on Wireless Communications},
  volume = {22},
  number = {9},
  pages = {5711--5724},
  doi = {10.1109/TWC.2023.3235910},
}

@article{hochwald2000systematic,
  title = {Systematic Design of Unitary Space-Time Constellations},
  author = {Hochwald, B.M. and Marzetta, T.L. and Richardson, T.J. and Sweldens, W. and Urbanke, R.},
  year = 2000,
  month = sep,
  journal = {IEEE Transactions on Information Theory},
  volume = {46},
  number = {6},
  pages = {1962--1973},
  doi = {10.1109/18.868472},
}

@article{hochwald2000unitary,
  title = {Unitary Space-Time Modulation for Multiple-Antenna Communications in {{Rayleigh}} Flat Fading},
  author = {Hochwald, B.M. and Marzetta, T.L.},
  year = 2000,
  month = mar,
  journal = {IEEE Transactions on Information Theory},
  volume = {46},
  number = {2},
  pages = {543--564},
  doi = {10.1109/18.825818},
}

@article{ishikawa2016subcarrierindex,
  title = {Subcarrier-Index Modulation Aided {{OFDM}} - Will It Work?},
  author = {Ishikawa, Naoki and Sugiura, Shinya and Hanzo, Lajos},
  year = 2016,
  journal = {IEEE Access},
  volume = {4},
  pages = {2580--2593},
  doi = {10.1109/ACCESS.2016.2568040}
}

@article{ishikawa201850,
  title = {50 Years of Permutation, Spatial and Index Modulation: From Classic {{RF}} to Visible Light Communications and Data Storage},
  author = {Ishikawa, Naoki and Sugiura, Shinya and Hanzo, Lajos},
  year = 2018,
  journal = {IEEE Communications Surveys \& Tutorials},
  volume = {20},
  number = {3},
  pages = {1905--1938},
  doi = {10.1109/COMST.2018.2815642}
}

@article{kammoun2003new,
  title = {A New Family of {{Grassmann}} Space-Time Codes for Non-Coherent {{MIMO}} Systems},
  author = {Kammoun, I. and Belfiore, J.-C.},
  year = 2003,
  month = nov,
  journal = {IEEE Communications Letters},
  volume = {7},
  number = {11},
  pages = {528--530},
  doi = {10.1109/LCOMM.2003.820081},
}

@article{edelman1999geometry,
author = {Edelman, Alan and Arias, Tom\'{a}s A. and Smith, Steven T.},
title = {The Geometry of Algorithms with Orthogonality Constraints},
year = {1999},
issue_date = {April 1999},
publisher = {Society for Industrial and Applied Mathematics},
address = {USA},
volume = {20},
number = {2},
doi = {10.1137/S0895479895290954},
journal = {SIAM J. Matrix Anal. Appl.},
month = apr,
pages = {303–353},
numpages = {51},
}

@article{kammoun2007noncoherent,
  title = {Non-Coherent Codes over the {{Grassmannian}}},
  author = {Kammoun, Ines and Cipriano, Antonio M. and Belfiore, Jean-Claude},
  year = 2007,
  month = oct,
  journal = {IEEE Transactions on Wireless Communications},
  volume = {6},
  number = {10},
  pages = {3657--3667},
  doi = {10.1109/TWC.2007.06059},
}

@article{kato2025maximizing,
author={Kato, Taiki and Iimori, Hiroki and Pradhan, Chandan and Malomsoky, Szabolcs and Ishikawa, Naoki},
  journal={IEEE Open Journal of the Communications Society}, 
  title={Maximizing Spectrum Efficiency of Data-Carrying Reference Signals via Bayesian Optimization}, 
  year={2025},
  volume={6},
  number={},
  pages={3892-3903},
  doi={10.1109/OJCOMS.2025.3562774}}

@article{konishi2022novel,
  title = {Novel {{MIMO}} Method for High Mobility Environments Using {{Schubert}} Cell Decomposition},
  author = {Konishi, Tatsumi and Nakano, Hiroyuki and Yano, Yoshikazu and Aoki, Michihiro},
  year = 2022,
  month = oct,
  journal = {IEICE Communications Express},
  volume = {11},
  number = {10},
  pages = {679--684},
  doi = {10.1587/comex.2022XBL0106}
}

@article{lee2016channel,
  title = {Channel Estimation via Orthogonal Matching Pursuit for Hybrid {{MIMO}} Systems in Millimeter Wave Communications},
  author = {Lee, Junho and Gil, Gye and Lee, Yong H.},
  year = 2016,
  month = jun,
  journal = {IEEE Transactions on Communications},
  volume = {64},
  number = {6},
  pages = {2370--2386},
  doi = {10.1109/TCOMM.2016.2557791},
}

@article{liu2015space,
  title = {Space Shift Keying for {{LOS}} Communication at {{mmWave}} Frequencies},
  author = {Liu, Peng and Springer, Andreas},
  year = 2015,
  month = apr,
  journal = {IEEE Wireless Communications Letters},
  volume = {4},
  number = {2},
  pages = {121--124},
  doi = {10.1109/LWC.2014.2381671},
}

@article{love2003grassmannian,
  title = {Grassmannian Beamforming for Multiple-Input Multiple-Output Wireless Systems},
  author = {Love, D.J. and Heath, R.W. and Strohmer, T.},
  year = 2003,
  month = oct,
  journal = {IEEE Transactions on Information Theory},
  volume = {49},
  number = {10},
  pages = {2735--2747},
  doi = {10.1109/TIT.2003.817466},
}

@article{love2005limited,
  title = {Limited Feedback Unitary Precoding for Spatial Multiplexing Systems},
  author = {Love, D.J. and Heath, R.W.},
  year = 2005,
  month = aug,
  journal = {IEEE Transactions on Information Theory},
  volume = {51},
  number = {8},
  pages = {2967--2976},
  doi = {10.1109/TIT.2005.850152},
}

@article{love2008overview,
  title = {An Overview of Limited Feedback in Wireless Communication Systems},
  author = {Love, David and Heath, Robert and N. Lau, Vincent and Gesbert, David and Rao, Bhaskar and Andrews, Matthew},
  year = 2008,
  month = oct,
  journal = {IEEE Journal on Selected Areas in Communications},
  volume = {26},
  number = {8},
  pages = {1341--1365},
  doi = {10.1109/JSAC.2008.081002},
}

@article{ngo2020cubesplit,
  title = {Cube-Split: A Structured {{Grassmannian}} Constellation for Non-Coherent {{SIMO}} Communications},
  author = {Ngo, Khachoang and Decurninge, Alexis and Guillaud, Maxime and Yang, Sheng},
  year = 2020,
  month = jun,
  journal={IEEE Transactions on Wireless Communications},
  volume={19},
  number={3},
  doi={10.1109/TWC.2019.2959781}
}

@article{ngo2022joint,
  title = {Joint Constellation Design for Noncoherent {MIMO} Multiple-Access Channels},
  author = {Ngo, Khachoang and Yang, Sheng and Guillaud, Maxime and Decurninge, Alexis},
  year = 2022,
  month = jan,
  journal = {IEEE Transactions on Information Theory},
  volume = {68},
  number = {11},
  pages = {7281--7305},
  doi = {10.1109/TIT.2022.3189254}
}

@article{ngo2025noncoherent,
  title = {Noncoherent {{MIMO}} Communications: Theoretical Foundation, Design Approaches, and Future Challenges},
  author = {Ngo, Khachoang and Cuevas, Diego and Gil, Ruben de Miguel and Baeza, Victor Monzon and Armada, Ana Garcia and Santamaria, Ignacio},
  year = 2025,
  month = may,
  journal = {arXiv:2505.23172},
  doi = {10.48550/arXiv.2505.23172}
}

@article{ochiai2001distribution,
  title = {On the Distribution of the Peak-to-Average Power Ratio in {{OFDM}} Signals},
  author = {Ochiai, H. and Imai, H.},
  year = 2001,
  month = feb,
  journal = {IEEE Transactions on Communications},
  volume = {49},
  number = {2},
  pages = {282--289},
  doi = {10.1109/26.905885},
}

@article{ochiai2012instantaneous,
  title = {On Instantaneous Power Distributions of Single-Carrier {FDMA} Signals},
  author = {Ochiai, Hideki},
  year = 2012,
  month = apr,
  journal = {IEEE Wireless Communications Letters},
  volume = {1},
  number = {2},
  pages = {73--76},
  doi = {10.1109/WCL.2012.012012.110254},
}

@article{ochiai2013statistical,
  title = {Statistical Distributions of Instantaneous Power and Peak-to-Average Power Ratio for Single-Carrier {{FDMA}} Systems},
  author = {Ochiai, Hideki},
  year = 2013,
  month = sep,
  journal = {Physical Communication},
  volume = {8},
  pages = {47--55},
  doi = {10.1016/j.phycom.2012.10.003}
}

@article{qin2018sparse,
  title = {Sparse Representation for Wireless Communications: A Compressive Sensing Approach},
  author = {Qin, Zhijin and Fan, Jiancun and Liu, Yuanwei and Gao, Yue and Li, Geoffrey Ye},
  year = 2018,
  month = may,
  journal = {IEEE Signal Processing Magazine},
  volume = {35},
  number = {3},
  pages = {40--58},
  doi = {10.1109/MSP.2018.2789521},
}

@article{taojiang2008overview,
  title = {An Overview: Peak-to-Average Power Ratio Reduction Techniques for {OFDM} Signals},
  author = {Jiang, Tao and Wu, Yiyan},
  year = 2008,
  month = jun,
  journal = {IEEE Transactions on Broadcasting},
  volume = {54},
  number = {2},
  pages = {257--268},
  doi = {10.1109/TBC.2008.915770},
}

@article{tse2005fundamentals,
    author    = {D. Tse and P. Viswanath},
    title     = {Fundamentals of Wireless Communication},
    journal = {Cambridge University Press},
    year = {2005}
}

@article{choi2017compressed,
  author={Choi, Jun Won and Shim, Byonghyo and Ding, Yacong and Rao, Bhaskar and Kim, Dong In},
  journal={IEEE Communications Surveys \& Tutorials}, 
  title={Compressed Sensing for Wireless Communications: Useful Tips and Tricks}, 
  year={2017},
  volume={19},
  number={3},
  pages={1527-1550},
  doi={10.1109/COMST.2017.2664421}}

@article{3gpp_ts_38211_v1820,
  title = {{{TS}} 138 211 - {{V18}}.2.0 - {{5G}}; {{NR}}; {{Physical}} Channels and Modulation ({{3GPP TS}} 38.211 Version 18.2.0 {{Release}} 18)},
  author = {{3GPP}},
  year = {2024}
}

@misc{ericsson2025ericsson,
    author = {Ericsson} ,
    title = {Ericsson Mobility Report},
    year = {2025}
}

@article{hansen2007schubert,
title = {Schubert unions in {Grassmann} varieties},
journal = {Finite Fields and Their Applications},
volume = {13},
number = {4},
pages = {738-750},
year = {2007},
doi = {https://doi.org/10.1016/j.ffa.2007.06.003},
author = {Johan P. Hansen and Trygve Johnsen and Kristian Ranestad},
}

@article{zhang20196g,
  title = {{6G} Wireless Networks: Vision, Requirements, Architecture, and Key Technologies},
  author = {Zhang, Zhengquan and Xiao, Yue and Ma, Zheng and Xiao, Ming and Ding, Zhiguo and Lei, Xianfu and Karagiannidis, George K. and Fan, Pingzhi},
  year = 2019,
  month = sep,
  journal = {IEEE Vehicular Technology Magazine},
  volume = {14},
  number = {3},
  pages = {28--41},
  doi = {10.1109/MVT.2019.2921208},
}

@article{ishikawa2017generalized,
  author={Ishikawa, Naoki and Rajashekar, Rakshith and Sugiura, Shinya and Hanzo, Lajos},
  journal={IEEE Transactions on Vehicular Technology}, 
  title={Generalized-Spatial-Modulation-Based Reduced-{RF}-Chain Millimeter-Wave Communications}, 
  year={2017},
  volume={66},
  number={1},
  pages={879-883},
  doi={10.1109/TVT.2016.2555378}}

@article{bell2008efficient,
    author = {Nathan Bell and Michael Garland},
    title =  {Efficient Sparse Matrix-Vector Multiplication on {CUDA}},
    journal = {NVIDIA Corporation},
    year = {2008},
}

@article{kincaid1989ITPACKV,
    author = {David R. Kincaid and Thomas C. Oppe and David M. Young},
    title = {{ITPACKV 2D User's Guide}},
    journal = {Center for Numerical Analysis, University of Texas at Austin},
    year = {1989},
    month = may
}

@article{yao2019semidefinite,
  author={Yao, Miao and Carrick, Matt and Sohul, Munawwar M. and Marojevic, Vuk and Patterson, Cameron D. and Reed, Jeffrey H.},
  journal={IEEE Transactions on Vehicular Technology}, 
  title={Semidefinite Relaxation-Based {PAPR}-Aware Precoding for Massive {MIMO-OFDM} Systems}, 
  year={2019},
  volume={68},
  number={3},
  pages={2229-2243},
  doi={10.1109/TVT.2018.2874819}}

@article{asano2026sparse,
    author = {Asano, Joe and Hama, Yuto and Iimori, Hiroki and Pradhan, Chandan and Malomsoky, Szabolcs and Ishikawa, Naoki},
    title = {Sparse {Grassmannian} Design for Noncoherent Codes via {Schubert} Cell Decomposition},
    year = {2026},
    journal = {arXiv:2601.21009},
    doi = {10.48550/arXiv.2601.21009}
}
}

\end{document}